\pdfoutput=1 %

\documentclass[conference,compsoc]{IEEEtran}

\usepackage{amsmath}
\usepackage[indLines=true,noEnd=true]{algpseudocodex}
\usepackage{booktabs}
\usepackage{float}
\usepackage{enumitem}
\setlist{nosep, leftmargin=*}
\usepackage{graphicx}
\usepackage[caption=false,font=footnotesize]{subfig}
\usepackage{xcolor}
\usepackage[varqu]{inconsolata}
\usepackage{listings}
\usepackage[
  backend=biber,
  style=numeric,
  sorting=none,
  maxbibnames=99,
  doi=false,
  isbn=false,
  url=false
]{biblatex}
\AtEveryBibitem{%
  \clearfield{pages}%
  \clearlist{publisher}%
  \clearlist{location}%
}
\DeclareSourcemap{
  \maps[datatype=bibtex]{
    \map[overwrite]{
      \step[fieldsource=series, final]
      \step[fieldset=booktitle, origfieldval]
      \step[fieldset=series, null]
    }
  }
}
\usepackage{tikz}
\usetikzlibrary{arrows.meta, positioning}
\usepackage{pgfplots}
\usepackage{pgfplotstable}
\pgfplotsset{compat=1.18}
\usepackage{balance}
\usepackage{eso-pic}
\usepackage{url}
\usepackage[hidelinks]{hyperref}
\AtBeginDocument{\urlstyle{tt}}
\hypersetup{
  pdftitle={PTSan: A Practical Memory Safety Sanitizer for C/C++ with Pointer-Object Authority},
  pdfauthor={Eli Davis, Eric Lahtinen, Michael Gordon},
  pdfsubject={Memory-safety sanitizers for C and C++},
  pdfkeywords={memory safety, sanitizer, pointer-based, object identity, tagged pointers, LLVM, spatial safety, temporal safety, C, C++}
}

\definecolor{codebackground}{RGB}{248,248,248}
\definecolor{codecomment}{RGB}{88,116,77}
\definecolor{codekeyword}{RGB}{36,92,150}
\definecolor{codestring}{RGB}{143,74,33}
\definecolor{codenumber}{RGB}{120,120,120}
\lstdefinestyle{ptsancode}{
  language=C,
  basicstyle=\ttfamily\scriptsize,
  keywordstyle=\color{codekeyword}\bfseries,
  commentstyle=\color{codecomment},
  stringstyle=\color{codestring},
  numbers=left,
  numberstyle=\scriptsize\color{codenumber},
  numbersep=6pt,
  stepnumber=1,
  backgroundcolor=\color{codebackground},
  frame=single,
  rulecolor=\color{black!20},
  columns=fullflexible,
  keepspaces=true,
  showstringspaces=false,
  breaklines=true,
  xleftmargin=1.4em,
  framexleftmargin=1.2em,
  aboveskip=0.3\baselineskip,
  belowskip=0.3\baselineskip
}

\providecommand{\Description}[1]{}

\begin{document}

\title{PTSan: A Practical Memory Safety Sanitizer for C/C++ \\ with Pointer-Object Authority}

\author{
  \IEEEauthorblockN{Eli Davis, Eric Lahtinen, and Michael Gordon}
  \IEEEauthorblockA{Aarno Labs\\
    Boston, MA, USA\\
    \texttt{\{\href{mailto:eli@aarno-labs.com}{eli},
    \href{mailto:elahtinen@aarno-labs.com}{elahtinen},
    \href{mailto:mgordon@aarno-labs.com}{mgordon}\}@aarno-labs.com}}
}

\maketitle

\AddToShipoutPictureFG*{%
  \AtPageLowerLeft{%
    \raisebox{0.5in}{\makebox[\paperwidth]{%
      \parbox{\textwidth}{\centering\footnotesize
        This work has been submitted to the IEEE for possible
        publication.\\ Copyright may be transferred without notice,
        after which this version may no longer be accessible.}}}%
  }%
}

\providecommand{\AblFullLamAll}{}\renewcommand{\AblFullLamAll}{46.4\%}
\providecommand{\AblLocalStripAll}{}\renewcommand{\AblLocalStripAll}{62.5\%}
\providecommand{\AblLocalStripPpLlvm}{}\renewcommand{\AblLocalStripPpLlvm}{5.9}
\providecommand{\AblLocalStripPpSpec}{}\renewcommand{\AblLocalStripPpSpec}{5.3}
\providecommand{\AblNoBoundsHoistAll}{}\renewcommand{\AblNoBoundsHoistAll}{112.0\%}
\providecommand{\AblNoCheckMergeAll}{}\renewcommand{\AblNoCheckMergeAll}{107.6\%}
\providecommand{\AblNoFilterAll}{}\renewcommand{\AblNoFilterAll}{109.2\%}
\providecommand{\AblNoLamAll}{}\renewcommand{\AblNoLamAll}{57.2\%}
\providecommand{\AblNoLoopHoistAll}{}\renewcommand{\AblNoLoopHoistAll}{86.5\%}
\providecommand{\AblNoLoopPreAll}{}\renewcommand{\AblNoLoopPreAll}{86.0\%}
\providecommand{\AblNoOptsAll}{}\renewcommand{\AblNoOptsAll}{139.9\%}
\providecommand{\AblNoStackIdAll}{}\renewcommand{\AblNoStackIdAll}{79.8\%}
\providecommand{\AblNoStripHoistAll}{}\renewcommand{\AblNoStripHoistAll}{66.2\%}
\providecommand{\AblStripHoistPpSpec}{}\renewcommand{\AblStripHoistPpSpec}{9}
\providecommand{\AsanOvAll}{}\renewcommand{\AsanOvAll}{117.4\%}
\providecommand{\LlvmIdFitFifteen}{}\renewcommand{\LlvmIdFitFifteen}{149}
\providecommand{\LlvmIdFitFifteenPct}{}\renewcommand{\LlvmIdFitFifteenPct}{89.2\%}
\providecommand{\LlvmIdFitSixteen}{}\renewcommand{\LlvmIdFitSixteen}{152}
\providecommand{\LlvmIdFitSixteenPct}{}\renewcommand{\LlvmIdFitSixteenPct}{91.0\%}
\providecommand{\LlvmMeasuredArm}{}\renewcommand{\LlvmMeasuredArm}{148}
\providecommand{\LlvmMeasuredLam}{}\renewcommand{\LlvmMeasuredLam}{149}
\providecommand{\LlvmMeasuredNolam}{}\renewcommand{\LlvmMeasuredNolam}{149}
\providecommand{\LlvmOvArmGeo}{}\renewcommand{\LlvmOvArmGeo}{40.8\%}
\providecommand{\LlvmOvArmMed}{}\renewcommand{\LlvmOvArmMed}{34.9\%}
\providecommand{\LlvmOvLamGeo}{}\renewcommand{\LlvmOvLamGeo}{27.2\%}
\providecommand{\LlvmOvLamMed}{}\renewcommand{\LlvmOvLamMed}{21.2\%}
\providecommand{\LlvmOvNolamGeo}{}\renewcommand{\LlvmOvNolamGeo}{31.5\%}
\providecommand{\LlvmOvNolamMed}{}\renewcommand{\LlvmOvNolamMed}{24.7\%}
\providecommand{\LlvmTestCount}{}\renewcommand{\LlvmTestCount}{167}
\providecommand{\LlvmUnderFiftyArm}{}\renewcommand{\LlvmUnderFiftyArm}{84}
\providecommand{\LlvmUnderFiftyArmPct}{}\renewcommand{\LlvmUnderFiftyArmPct}{57\%}
\providecommand{\LlvmUnderFiftyLam}{}\renewcommand{\LlvmUnderFiftyLam}{114}
\providecommand{\LlvmUnderFiftyLamPct}{}\renewcommand{\LlvmUnderFiftyLamPct}{77\%}
\providecommand{\LlvmUnderFiftyNolam}{}\renewcommand{\LlvmUnderFiftyNolam}{103}
\providecommand{\LlvmUnderFiftyNolamPct}{}\renewcommand{\LlvmUnderFiftyNolamPct}{69\%}
\providecommand{\MemAsanAll}{}\renewcommand{\MemAsanAll}{3.33$\times$}
\providecommand{\MemIdFitCount}{}\renewcommand{\MemIdFitCount}{8}
\providecommand{\MemPtsanAll}{}\renewcommand{\MemPtsanAll}{1.014$\times$}
\providecommand{\MemPtsanIdFit}{}\renewcommand{\MemPtsanIdFit}{1.021$\times$}
\providecommand{\MemPtsanMax}{}\renewcommand{\MemPtsanMax}{1.121$\times$}
\providecommand{\MemRsanAll}{}\renewcommand{\MemRsanAll}{3.01$\times$}
\providecommand{\MemRsanIdFit}{}\renewcommand{\MemRsanIdFit}{2.37$\times$}
\providecommand{\RsanOvAll}{}\renewcommand{\RsanOvAll}{50.6\%}
\providecommand{\ServerMaxId}{}\renewcommand{\ServerMaxId}{16{,}411}
\providecommand{\SpecBenchCount}{}\renewcommand{\SpecBenchCount}{17}
\providecommand{\SpecCommonRsanCount}{}\renewcommand{\SpecCommonRsanCount}{13}
\providecommand{\SpecIdExceedCount}{}\renewcommand{\SpecIdExceedCount}{7}
\providecommand{\SpecIdFitCount}{}\renewcommand{\SpecIdFitCount}{10}
\providecommand{\SpecOvArmAll}{}\renewcommand{\SpecOvArmAll}{54.7\%}
\providecommand{\SpecOvArmCommonRsan}{}\renewcommand{\SpecOvArmCommonRsan}{48.4\%}
\providecommand{\SpecOvArmFp}{}\renewcommand{\SpecOvArmFp}{43.5\%}
\providecommand{\SpecOvArmInt}{}\renewcommand{\SpecOvArmInt}{65.4\%}
\providecommand{\SpecOvLamAll}{}\renewcommand{\SpecOvLamAll}{46.4\%}
\providecommand{\SpecOvLamCommonRsan}{}\renewcommand{\SpecOvLamCommonRsan}{43.4\%}
\providecommand{\SpecOvLamFp}{}\renewcommand{\SpecOvLamFp}{37.3\%}
\providecommand{\SpecOvLamInt}{}\renewcommand{\SpecOvLamInt}{55.0\%}
\providecommand{\SpecOvNolamAll}{}\renewcommand{\SpecOvNolamAll}{57.2\%}
\providecommand{\SpecOvNolamCommonRsan}{}\renewcommand{\SpecOvNolamCommonRsan}{52.3\%}
\providecommand{\SpecOvNolamFp}{}\renewcommand{\SpecOvNolamFp}{43.3\%}
\providecommand{\SpecOvNolamInt}{}\renewcommand{\SpecOvNolamInt}{70.6\%}

\begin{abstract}

Memory safety errors remain the dominant source of severe
vulnerabilities in C and C++. Pointer-based sanitizers
provide stronger guarantees than location-based tools such as LLVM's
ASan, but their overhead and compatibility limitations have
constrained production use. We present PTSan, an LLVM sanitizer that
makes pointer-based checking practical by storing an object identifier
in each pointer's high bits and its bounds in a fixed-size runtime
table. This representation trades a finite live-object budget for low
overhead, optimizer visibility, and commodity-hardware
support. Because identity travels with the pointer value, ordinary
LLVM dataflow propagates it without explicit per-pointer metadata
instructions. Separating metadata lookup from check enforcement
exposes both as LLVM IR, enabling check hoisting, elision, and
merging, plus whole-function min-cut placement of compatibility tag
strips. Intel Linear Address Masking eliminates the remaining strips
in hardware when available.

On stock hardware PTSan runs at parity with the fastest published
location-based sanitizer: \SpecOvNolamAll{} geomean overhead on SPEC
CPU 2017 on x86-64 (\SpecOvLamAll{} with Intel LAM), \SpecOvArmAll{}
on ARM64, and \LlvmOvNolamGeo{} (x86-64) on the application-shaped
LLVM MultiSource suite.  This is roughly a third of the published
overhead percentage of prior systems with similar pointer-object
authority guarantees, while preserving the inter-object, non-object,
and temporal detection coverage measured by an independent memory
safety test suite. Its physical-memory overhead is effectively native,
a critical property for production deployment as memory costs become a
first-order constraint. We also demonstrate practical overhead on
real-world server and security workloads. These results show that
PTSan brings practical pointer-based memory safety into a
recompile-only sanitizer deployment model.

\end{abstract}

\IEEEpeerreviewmaketitle

\section{Introduction}
\label{sec:introduction}

Memory safety remains the central security problem of deployed
systems software.  C and C++ still implement operating systems,
browsers, language runtimes, databases, and performance-sensitive
servers, and the manual bounds and lifetime discipline they require
routinely fails: memory-safety errors account for the majority of
severe vulnerabilities reported by major
vendors~\cite{cisa-memory-safety}, including about 70\% of Chromium's
high-severity bugs~\cite{chromium-memory-safety}, and they are the
substrate for code-reuse attacks, data-only attacks, information
disclosure, and remote
compromise~\cite{szekeres2013eternal,song2019sok}.  The problem is
accelerating: LLM agents already make exploiting known
vulnerabilities faster and cheaper~\cite{fang2024oneday}, and they
are beginning to find and exploit previously unknown, unpatched
ones~\cite{fang2024zeroday}.  Because these codebases cannot be
rewritten in memory-safe languages quickly, they need protection
that arrives before each individual bug is found, triaged, and
patched.

Sanitizers are the natural vehicle for that protection, adding
memory-safety guarantees by recompilation, without source changes.
Redzone sanitizers, the most popular category, are ABI compatible, and the
model is proven: AddressSanitizer (ASan) is used throughout industry
to catch memory errors~\cite{serebryany2012asan}. But their own
authors position them as testing tools, not deployed defenses, and not
by accident: they attach metadata to memory locations and pay for that
compatibility twice. First in guarantee: ASan and the faster
RangeSanitizer (RSan)~\cite{gorter2025rangesanitizer} check whether an
address is valid, not whether the pointer is entitled to access it, so
an access redirected into another live object passes their checks,
exactly the freedom an adaptive attacker needs. Second in memory:
redzones, shadow memory, and quarantine triple physical memory use
(3.33$\times$ for ASan and 3.01$\times$ for RSan on SPEC CPU 2017),
disqualifying them in production fleets where memory is a first-order
cost. Pointer-based systems close the guarantee gap by binding each
pointer to the object it may access, but at costs that have kept them
out of deployment: SoftBound+CETS breaks the ABI by passing pointer
bounds as extra function arguments, and runs at 161\% overhead with
2.7$\times$ memory~\cite{10.1145/3642974.3652285}; CUP costs
158\%~\cite{burow2018cup}.

We present PTSan, an LLVM sanitizer that provides pointer-based
checking at redzone-sanitizer speed and near-native memory.  PTSan
stores a compact object identifier in the high bits of each pointer;
the identifier indexes a flat runtime table holding the object's base
and span.  Before a memory access, PTSan checks that the entire
access range lies within the bounds of the object the pointer names.
When an object is freed, its entry is cleared.  Once identifiers are
reused, a stale pointer is accepted only if a replacement object at
its address receives the same identifier, providing probabilistic
temporal protection.

Two properties make PTSan fast where
prior pointer-based systems were slow.  Identity propagation is free:
the identifier lives in the pointer value, so ordinary dataflow
(copies, arithmetic, pointer loads and stores) carries it
without metadata instructions.  And the work that
remains at an access stays ordinary LLVM IR, with bounds loads kept
separable from the checks they feed, so the compiler can hoist
identifier extraction and bounds loads, merge and elide checks, and
replace per-iteration loop checks with whole-loop range checks
computed before LLVM's loop optimizations restructure the code.

PTSan's remaining design decision is a deliberate compatibility
tradeoff: it supports $2^{16}$ simultaneously live object identifiers.
The default width follows from ordinary x86-64 allocation:
\texttt{glibc}'s \texttt{malloc} returns addresses in the low 48-bit
user address space, leaving 16 pointer bits in which PTSan can carry
object identity.  PTSan can therefore retain the system allocator
unchanged rather than requiring a custom allocator or partitioned
address space.  The compact identifier also bounds metadata to roughly
a fixed table of 1.2\,MiB, with no redzones, shadow memory, or
quarantine.  Support for Intel's Linear Address Masking
(LAM)~\cite{intel-lam} is a further benefit of the design: LAM\_U48
masks 15 identifier bits during memory accesses, removing most
software tag stripping while reducing the ID budget to $2^{15}$.  The
cost is a finite live-object limit, but the budget covers most
programs we measured: \LlvmIdFitSixteenPct{} of the \LlvmTestCount{}
LLVM MultiSource programs fit the default configuration.  Deployments
can measure peak demand in advance, and strict mode fails closed if
the budget is exceeded.

Our overhead results\footnote{We report full-suite overheads including
ID-exhausting workloads. These runs fall outside PTSan's
strong-guarantee regime, but ID propagation and access checking remain
active. We retain them so the aggregates capture instrumentation cost
across the full evaluated suite.} move pointer-based memory safety
into a new performance regime. On SPEC CPU 2017, PTSan's geomean
runtime overhead is \SpecOvNolamAll{} on stock x86-64 hardware: parity
with RSan, the fastest published redzone sanitizer, roughly half of
ASan's overhead on the same machine, and a third of the published
overhead percentage of prior pointer-based systems.  When Intel LAM
elides the tag stripping that remains after the compiler has hoisted
most of it, the geomean falls to \SpecOvLamAll{}.  The result is
portable: recompiling for ARM64 yields \SpecOvArmAll{} on the same
suite, with no hardware assist.  SPEC's pointer-dense compute kernels
are a stress test rather than PTSan's target workload; on the broader
149-program LLVM MultiSource suite the geomean is \LlvmOvNolamGeo{} on
x86-64 (\LlvmOvLamGeo{} with LAM, \LlvmOvArmGeo{} on ARM64), and seven
real-world server and cryptographic workloads (Redis, PostgreSQL,
NGINX, OpenSSL, memcached, Apache, SQLite) add overheads ranging from
near-native to 44\% on x86-64, each well within the $2^{16}$
identifier budget (peak demand \ServerMaxId{}).

PTSan's physical-memory overhead is largely fixed and measured at
\MemPtsanAll{} geomean with a worst case of \MemPtsanMax{} on SPEC.
PTSan is effectively native, where ASan costs 3.33$\times$ and RSan
3.01$\times$ on the same benchmarks and baseline, and SoftBound+CETS
self-reports the cost at 2.7$\times$~\cite{10.1145/3642974.3652285}.
For production deployment this is the decisive number: PTSan's
footprint is bounded by design and measured near-native, at the cost
of only a slice of applicability.

PTSan's strong safety guarantee is also measurable.  On
MSET~\cite{vintila2025mset}, an independent benchmark suite for
assessing memory-safety sanitizers, PTSan matches the originally
published coverage of SoftBound+CETS, the strongest prior
pointer-based system, detecting all 126 inter-object spatial and all
40 temporal cases; it is the only low-overhead system that detects
non-linear out-of-bounds accesses that land inside another live
object, an access pattern available in real-world vulnerabilities.

This paper makes the following contributions:

\noindent\textbf{Design.} A sanitizer design that carries a compact
object identifier in the pointer and keeps bounds in a small fixed
table.  This design makes identity propagation free, checks ordinary
optimizable LLVM IR, and memory overhead bounded by construction.  We
present a correctness argument covering both the base instrumentation
and its optimizations.

\noindent\textbf{Optimizations.}  Compiler analyses that exploit the
separability of bounds loads from checks: identifier-extraction and
bounds-load hoisting, range-check merging and elision, loop
preanalysis whose results survive LLVM's loop transformations, and a
min-cut placement of tag strips and retags that hoists stripping out
of hot loops.

\noindent\textbf{Implementation.}  A complete LLVM/compiler-rt
sanitizer for x86-64 and AArch64 covering stack, heap, global, and
mmap allocation, libc interposition, and software or LAM-accelerated
identifier stripping, released as open source at
\url{https://github.com/Aarno-Labs/PointerTagSanitizer}.

\noindent\textbf{Evaluation.}  Across SPEC CPU 2017, 149 LLVM
MultiSource programs, MSET, and real server workloads, on two
architectures: runtime overhead at parity with the fastest redzone
sanitizer, near native memory overhead, inter-object and temporal
detection coverage matching the strongest pointer-based
system, and identifier demand that validates the bounded-ID design for
its target class of programs.

\section{Background}
\label{sec:background}

\subsubsection*{Memory-Safety Errors and Checking Models}
Memory-safety violations are usually divided into spatial and temporal
errors.  A \emph{spatial} error occurs when a program accesses memory
outside the object that a pointer is allowed to reference.  Following
MSET, we further distinguish where the illegal access lands: an
\emph{inter-object} access lands inside a different currently allocated
object, a \emph{non-object} access lands in unallocated or padding
memory, and an \emph{intra-object} access stays within its allocation
but crosses an internal field or subobject
boundary~\cite{vintila2025mset}.  A \emph{temporal} error occurs when a
pointer is used after the lifetime of its object has ended, or when a
deallocation operation does not match the object being deallocated.  In
this paper, unless stated otherwise, \emph{object} means an allocation
object: a heap allocation, stack allocation, global object, or mmap
allocation.  PTSan enforces bounds at this granularity and does not
claim intra-object field protection
(Section~\ref{sec:security:limitations}).

Sanitizers for these errors follow one of two checking models.  A
\emph{location-based} sanitizer attaches validity metadata to memory
addresses and asks, at each access, ``is this target address valid?'';
ASan's poisoned redzones are the canonical
example~\cite{serebryany2012asan}.  A \emph{pointer-based} sanitizer
associates each pointer with the object from which it was derived and
asks, ``is this pointer allowed to access this
range?''; SoftBound+CETS is the canonical
example~\cite{nagarakatte2009softbound,nagarakatte2010cets}.  The
models differ exactly when an illegal access lands in valid memory: a
location-based check accepts an access that escapes its object and
lands inside another live object, while a pointer-based check rejects
it, because the authority carried by the pointer still names the
original object.  PTSan is a pointer-based sanitizer, and such
inter-object errors arise in real vulnerabilities, as the next
subsection illustrates.

\subsubsection*{A Recent Inter-Object Overflow}
\label{sec:background:cve-example}
CVE-2025-57807, a critical (CVSS~9.8) heap overflow in ImageMagick
7.1.2-0, shows why this distinction is security-relevant in current C
code~\cite{imagemagick2025cve57807}.  A \texttt{BlobStream} forward
seek advances the write offset past the buffer's capacity; the next
write grows the buffer by the write length rather than by the full
distance to the offset, then forms its destination by adding the
attacker-controlled offset to the buffer base.  The write therefore
lands wherever the offset points, not merely past the end of the
source buffer.  Its ingredients are mundane: an offset or index an
attacker can influence, and a resize that fails to account for it, a
pattern that recurs throughout buffer, stream, and serialization
code.  When the offset is chosen to land inside a different live
allocation, the destination is valid, mapped memory, and address- or
range-validity checks have nothing to reject.

Building a working exploit was straightforward.  A small harness
drives the vulnerable \texttt{WriteBlob} path directly and steers the
write into a second live heap object; reproducing it under ASan took
only modest tuning, an oversized backing allocation and a predicted
\texttt{realloc} relocation offset to absorb ASan's allocator
perturbation.  Both ASan and RSan~\cite{gorter2025rangesanitizer}
then accept the corrupting write, since its destination is an
allocated, address-valid object, while PTSan detects it because the
writing pointer's object ID authorizes the source buffer, not the
object the write lands in.

\section{Threat Model}
\label{sec:threat-model}

This section states the assumptions under which PTSan makes its security
claims and the attacker it defends against.  Once the mechanism is defined
(Section~\ref{sec:design}), Section~\ref{sec:base-correctness} gives the
security analysis.

\subsection{Assumptions}
\label{sec:security:assumptions}

We assume the toolchain is correct and the instrumentation is complete:
the compiler, the C library, and PTSan itself implement their
specifications, and the relevant allocations, deallocations, and memory
accesses are instrumented or intercepted.  Completeness is the practical
limit of a young sanitizer rather than a property of the design;
uninstrumented allocation sites and unsupported operations are discussed
in Section~\ref{sec:security:coverage}.

The remaining assumptions are structural.  First, the program has at
most $2^{16}$ live objects at any one time, or $2^{15}$ when Linear Address
Masking is enabled, so every live object can hold a distinct identifier;
in strict mode, identifier exhaustion fails closed.  Second, program
addresses fit in the low 48 bits, leaving the high 16 bits for the
identifier (Section~\ref{sec:design:pointer-representation}).
Third, address-space layout randomization (ASLR) is enabled.  For
temporal claims we additionally assume the program is free of data races
between freeing an object in one thread and accessing it in another.

We distinguish strict and permissive modes.  Strict mode is the
configuration for the security claims in this paper: identifier
exhaustion and accesses through the untagged (zero) or default
identifier fail closed.  Permissive mode keeps such accesses running and
is an engineering aid for bring-up and compatibility debugging, not a
security configuration.

\subsection{Attacker Model}
\label{sec:threat-model:attackers}

We assume an attacker who can trigger memory-safety bugs in
instrumented C/C++ code by controlling program inputs: offsets,
lengths, indices, object contents, allocation sizes, and control flow
along vulnerable paths.  Two independent axes describe such an
attacker: how much of the system the attacker knows and controls, and
what bug primitive the vulnerability gives them.  The security
analysis (Section~\ref{sec:base-correctness}) walks this grid.

\subsubsection{Knowledge and environment control}

We distinguish three attacker levels, from strongest to weakest:

\begin{itemize}[leftmargin=*]
  \item \textbf{A1 (omniscient, controls all allocation).}  The attacker
  understands PTSan and the program perfectly and can predict exactly which
  identifier each allocation receives.  This is the worst case for
  PTSan, and it is also the least realistic: it requires the attacker to
  be able to model all allocations throughout the program's life,
  which is difficult for a program serving other requests.

  \item \textbf{A2 (omniscient, cannot control allocation).}  The
  attacker understands PTSan and the program but there is enough noise in
  the system that they cannot predict identifier
  assignment.  This is the realistic strong attacker.

  \item \textbf{A3 (understands the program, not PTSan).}  The attacker
  understands the target program but is not aware of the specific defense.
\end{itemize}

\subsubsection{Bug primitives}  Orthogonally, we distinguish what the
vulnerability lets any of these attackers do:

\begin{itemize}[leftmargin=*]
  \item \textbf{P1 (length control).}  The attacker controls the length
  of a contiguous access through a legitimately derived pointer, such
  as a length passed to \texttt{memcpy} or the trip count of a loop
  that advances element by element.  The accessed range grows outward
  from the source object and cannot skip over adjacent memory.

  \item \textbf{P2 (offset control).}  The attacker controls an index,
  offset, or stride combined with a legitimately derived pointer, so
  the resulting access can land at a discontiguous address, including
  inside another live object.  This is the shape of the non-linear
  out-of-bounds errors in MSET's taxonomy and of the ImageMagick
  vulnerability of Section~\ref{sec:background:cve-example}.

  \item \textbf{P3 (pointer forgery).}  The attacker constructs full
  pointer values, including the high bits, through integer-to-pointer
  manipulation or arithmetic with very large numbers.  A fabricated
  value did not obtain its identifier from
  an allocation, so the dataflow argument for checked accesses no
  longer applies, and we analyze this primitive separately
  (Section~\ref{sec:security:forgery}).

  \item \textbf{P4 (temporal primitives).}  The attacker holds a stale
  pointer to a freed object and can influence when the program
  allocates and frees objects.
\end{itemize}

PTSan's primary security property is object-bounds enforcement for
checked accesses: a memory access through a tagged pointer must lie
within the live allocation object named by the pointer's identifier.
The next sections define the mechanism and then analyze what each
attacker level achieves with each primitive.

\section{PTSan Design}
\label{sec:design}

PTSan is a recompile-only LLVM sanitizer for existing 64-bit C/C++
programs whose addresses fit in the low 48 bits: protected code is
compiled with PTSan instrumentation, linked against the PTSan
runtime, and run unchanged.  It is built around a simple
representation: each protected pointer
carries the identity of the allocation object from which it was
derived.  The runtime stores bounds and lifetime state for that object
in compact tables indexed by the ID.  A memory access therefore checks
the object named by the pointer, rather than asking only whether the
target address is currently mapped or surrounded by poisoned memory.

\subsection{Tagged Pointer Representation}
\label{sec:design:pointer-representation}

\begin{figure}[t]
  \centering
  \begin{tikzpicture}[font=\scriptsize, >={Latex[length=1.6mm]}]
    \draw[fill=black!12] (0,0) rectangle (1.3,0.36);
    \draw[fill=black!3] (1.3,0) rectangle (4.3,0.36);
    \node at (0.65,0.18) {object ID};
    \node at (2.8,0.18) {canonical address};
    \node[font=\tiny, anchor=south west] at (-0.05,0.37) {63};
    \node[font=\tiny, anchor=south east] at (1.28,0.37) {48};
    \node[font=\tiny, anchor=south west] at (1.32,0.37) {47};
    \node[font=\tiny, anchor=south east] at (4.35,0.37) {0};
    \node[font=\tiny, anchor=south west, inner sep=1pt] at (4.9,-0.16) {bounds table};
    \draw[fill=black!3] (4.9,-0.44) rectangle (8.1,-0.20);
    \node[font=\tiny] at (6.5,-0.32) {$\cdots\cdots$};
    \draw[fill=black!12] (4.9,-0.76) rectangle (8.1,-0.44);
    \draw (6.5,-0.76) -- (6.5,-0.44);
    \node at (5.7,-0.60) {base};
    \node at (7.3,-0.60) {span};
    \draw[fill=black!3] (4.9,-1.00) rectangle (8.1,-0.76);
    \node[font=\tiny] at (6.5,-0.88) {$\cdots\cdots$};
    \draw[->, thick] (0.65,0) |- (4.9,-0.60);
  \end{tikzpicture}
  \caption{PTSan pointer layout: the 16-bit object ID indexes a flat
  table of the object's base and span; the low 48 bits remain
  the canonical address.}
  \Description{A 64-bit pointer word with the object ID in bits 63
  down to 48 and the canonical address in bits 47 down to 0.  An
  arrow from the ID field indexes an entry of a flat bounds table;
  the entry holds the object's base and span.}
  \label{fig:pointer-layout}
\end{figure}
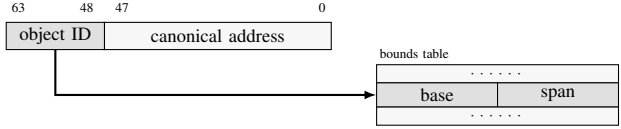

PTSan uses the upper 16 bits of a 64-bit pointer as an object ID and
the lower 48 bits as the canonical program address
(Figure~\ref{fig:pointer-layout}).  The
implementation uses two reserved
identifiers.  ID 0 names untagged pointers.  The highest ID is a
default ID used by permissive mode when the runtime cannot assign a
normal object ID.  The remaining IDs are assigned to protected stack,
heap, mmap, and global objects.  In strict mode, untagged and
default-ID pointers are invalid, allowing for the strong security guarantee; in
permissive mode, they are open as a compatibility fallback.

Embedding the ID in the pointer is the main reason PTSan avoids a
separate metadata-propagation problem.  LLVM copies, PHI nodes, select
instructions, bitcasts, loads and stores of pointer values, and
derived pointer computations move the full pointer value.  Because the
ID is part of that value, ordinary dataflow carries it along with the
address.  A pointer derived from an allocation therefore continues to
name the same allocation even when its low address changes through
pointer arithmetic, and the work that remains at a dereference
(identifier extraction, bounds loads, and range checks) stays
ordinary LLVM IR that the compiler can optimize
(Section~\ref{sec:optimizations}).

When a raw address is required, PTSan removes the high bits before the
address is observed.  The runtime provides an ID-stripping pass and
call-boundary handling for operations that must receive canonical
pointers.  By default this strip is a software mask inserted by the
compiler and runtime; Section~\ref{sec:optimizations:strip-hoisting}
describes how the compiler places these strips to minimize how often
they execute.  On systems with Intel Linear Address Masking
(LAM)~\cite{intel-lam}, the hardware ignores the tag bits during address translation, so
PTSan can carry the identifier through most memory operations without an
explicit per-access strip. The compiler and runtime still strip at
boundaries where software expects canonical addresses, such as pointers
passed to the kernel.  PTSan uses the LAM\_U48 variant, which masks
bits 62--48, so the LAM configuration uses
a 15-bit identifier: bit~63 is reserved by the hardware mask, and the
live-object budget is correspondingly smaller.  Because ordinary
x86-64 program addresses occupy only the low 48
bits~\cite{linux-x86-64-mm}, masking these bits leaves the canonical
address intact.  On AArch64, PTSan
uses the full 16-bit identifier and strips in software throughout:
the architecture's Top-Byte-Ignore feature masks only the top eight
bits, too few for PTSan's identifier, so PTSan does not use it.  We
report all three configurations (x86-64, x86-64 with LAM, and ARM64)
in Section~\ref{sec:evaluation}.

\subsection{Runtime Metadata}
\label{sec:design:metadata}

PTSan stores object bounds in a fixed-size runtime table.  Each ID
indexes a bounds entry containing two 64-bit fields: the first has the
ID in its upper 16 bits and the base address in its lower 48
(\texttt{base}), and the second is the size of the allocation
(\texttt{span}).  Metadata retrieval is therefore a direct index into
a flat table, with no shadow-memory walk or multi-level lookup.  The
runtime also maintains ID-pool state for non-stack objects and
thread-local state for stack IDs.

The bounds entry is the authoritative metadata for an object.  When an
object is live, its entry contains the above struct.  When the object
is no longer live, PTSan clears the entry, setting base and span to
zero.  This single transition is used by both spatial and temporal
checks: spatial checks compare an access range against the live
bounds, while immediate stale dereferences fail because the span is
zero.

\subsection{Allocation Instrumentation}
\label{sec:design:allocation}

PTSan assigns IDs at the points where allocation objects enter the
protected program.  For heap and mmap objects, sanitizer interceptors
wrap allocation APIs such as \texttt{malloc}, \texttt{calloc},
\texttt{realloc}, and \texttt{mmap}.  On allocation, the wrapper
obtains an ID, writes the object's base and span into the metadata
table, and returns the pointer with the ID in the upper bits.  If an
operation changes an allocation's size, the runtime updates the span
for the object's ID.  By default, the recorded span is the requested allocation
size rather than the allocator's usable size, so PTSan also rejects
accesses that stray into allocator slack beyond the request.

For stack objects, the LLVM pass instruments eligible
\texttt{alloca} instructions.  At function entry, PTSan obtains the
IDs needed for the function's stack allocations and writes their
bounds.  The common case uses thread-local contiguous ranges of stack
IDs, avoiding a runtime call for each function that contains at least
one \texttt{alloca}.
If the fast path
cannot supply the requested IDs, the pass calls a runtime helper that
allocates and initializes the IDs.  At function returns and other exits,
including unwind paths, the instrumentation returns the IDs to the
runtime.  In PTSan's stack-temporal mode, it also clears the
corresponding bounds entries and batches pending stack-ID returns so
that consecutive calls to the same function do not immediately reuse
the same stack IDs; delaying reuse this way makes stack-buffer-leak
detection more robust.

Globals are handled by introducing tagged global-pointer proxies.  At
startup, a PTSan constructor assigns IDs to instrumented globals,
writes their bounds, creates tagged pointer values, and stores those
values in the proxies.  Uses of the original global in instrumented
code are redirected through the proxy, so derived pointers carry the
global's ID in the same way as heap and stack pointers.  External or
unsupported globals can be represented conservatively, but the
object-bounds guarantee of Section~\ref{sec:base-correctness} applies
to globals for which PTSan creates precise IDs and bounds.

\subsection{Access Checks}
\label{sec:design:checks}

PTSan instruments LLVM memory operations with range checks.  The pass
identifies pointer operands of scalar loads and stores, contiguous
vector loads and stores, supported atomic operations,
compare-and-exchange operations, and memory intrinsics such as
\texttt{memcpy}, \texttt{memmove}, and \texttt{memset}.  For
fixed-width operations, the access size comes from the LLVM type.  For
memory intrinsics, the size operand defines the checked range, so a
variable-length copy or set requires one range check per pointer
operand rather than one check per byte.

For a pointer \texttt{ptr} and access size \texttt{size}, the base
check is:

\begin{lstlisting}
id     = ptr >> 48;
base   = ptsan_bounds[id].base;
span   = ptsan_bounds[id].span;

ok = (size <= span) && (ptr - base <= span - size);
\end{lstlisting}

This check validates an access range, not only the first accessed
address.  The first comparison ensures that the requested access can
fit within the object at all.  The second comparison verifies that the
low address is no earlier than the object's base and that the range
ending at \texttt{ptr + size} does not exceed the object bound.  As a
result, PTSan detects both ordinary out-of-bounds offsets and
type-width errors in which a pointer to a small object is used for a
larger access.

On failure, the inserted code branches to a reporting path that calls
the PTSan runtime with the check site, pointer ID, access type, address,
and size.  The passing path is straight-line integer arithmetic,
metadata loads, and conditional branches.  This is the mechanism that
later optimizations operate on: the pass can eliminate or move checks,
but the base check defines the condition that an access must satisfy.

\subsection{Deallocation and Temporal Enforcement}
\label{sec:design:deallocation}

Deallocation is the inverse of allocation.  A free-like operation
extracts the ID from the pointer and checks that the ID names a live
object whose recorded base address matches the pointer.  If it does
not, PTSan reports an invalid or double free.  If it does, PTSan clears
the bounds entry and returns the ID to the appropriate pool.

Clearing bounds means a stale nonzero-size dereference fails
deterministically until the ID is reused.  A stale pointer still
carries its old ID, but that ID's span is zero, so the range check
fails.  After ID reuse, temporal protection is probabilistic.  A stale
pointer can pass only if the reused ID now describes a live object
whose base address matches it.  This still requires the stale pointer
to line up with both the reused ID and the new object's address, but
it is not a deterministic temporal safety guarantee.
Section~\ref{sec:security:temporal} analyzes how hard that coincidence
is to arrange: for an attacker who cannot predict identifier assignment
lining up reuse and overlap succeeds only with very low
probability.

\subsection{Interoperability Boundaries}
\label{sec:design:interoperability}

PTSan is designed for deployment on existing C/C++ codebases, so it has
to coexist with code that expects ordinary pointers.  The compiler pass
strips IDs before operations and calls that require canonical
addresses.  The runtime also tracks when execution is inside libc so
that internal allocator and libc activity does not accidentally create
application-visible tags.

The important distinction is between security boundaries and
compatibility boundaries.  Allocation and deallocation wrappers are
security-critical because they create, resize, and retire object
metadata.  Generic libc wrappers and ID stripping are compatibility
mechanisms because they prevent high-bit tags from leaking into code
that cannot interpret them.  If uninstrumented code performs a memory
access after receiving a stripped pointer, that access is not covered by
PTSan's core check (Section~\ref{sec:security:coverage}).

\section{Security Analysis}
\label{sec:base-correctness}
\label{sec:security}
\label{sec:guarantees:model}

\begin{table}[t]
  \centering
  \caption{Outcomes per attacker level and bug primitive.}
  \label{tab:attacker-outcomes}
  \footnotesize
  \resizebox{\columnwidth}{!}{%
    \begin{tabular}{@{}llll@{}}
      \toprule
      Primitive & A1 & A2 & A3 \\
      \midrule
      P1: length            & blocked & blocked & blocked \\
      P2: offset            & blocked & blocked & blocked \\
      P3: forgery, heap     & blocked under ASLR & blocked under ASLR &
        blocked under ASLR \\
      P3: forgery, stack    & stack variables only$^\dagger$ &
        stack variables only$^\dagger$ & blocked \\
      P4: temporal primitives  & reuse grooming &
        $\sim 2^{-16}$/attempt & $\sim 2^{-16}$/attempt \\
      \bottomrule
    \end{tabular}%
  }

  \smallskip
  {\footnotesize\raggedright $^\dagger$Confined to user-declared
  variables; saved return addresses and other system data are unreachable
  (Section~\ref{sec:security:forgery}).\par}
\end{table}

The base guarantee is for checked LLVM IR memory operations.  Assuming
that allocation and deallocation events are instrumented or
intercepted, that every in-scope memory operation is instrumented,
that strict ID mode is used, and that high pointer bits are not forged
by explicit integer-pointer manipulation, every checked nonzero-size
access through a PTSan pointer either lies within the live allocation
object named by that pointer's ID or traps before the access executes.
Table~\ref{tab:attacker-outcomes} summarizes the outcomes.

The claim is object-granular.  PTSan tracks heap allocations, stack
allocations, globals, and mmap regions as allocation objects.  It does
not claim subobject safety.  The claim also applies only at
instrumented or intercepted boundaries: stripped-pointer accesses in
fully uninstrumented code are outside the core guarantee.

\subsection{Runtime Invariants}
\label{sec:guarantees:invariants}

The base argument relies on four invariants that map directly to PTSan's
runtime metadata and instrumentation.

\textbf{Allocation metadata.}  For each live tracked object, the runtime
records a nonzero object ID, a low-address base, and a span.  The
bounds entry for that ID describes exactly the
allocation object for which the ID was assigned.

\textbf{Live-ID uniqueness.}  A normal ID names at most
one live object at a time. In strict mode, ID 0 and the default ID do not authorize
ordinary accesses.  When no normal ID is available, the allocation
receives the default ID, which carries no bounds, so any access through
it traps; strict mode therefore fails closed rather than granting broad
bounds.

\textbf{Lifetime invalidation.}  When a tracked object's lifetime ends,
PTSan clears the object's bounds entry before the ID can authorize later
accesses.  Heap and mmap deallocation paths also check that the pointer
being deallocated names the recorded base address before clearing
metadata.

\textbf{Dereference mediation.}  In the unoptimized transformation,
every protected LLVM memory operation is preceded by a check over the
same pointer authority and an access range that covers the operation.
This includes scalar loads and stores, supported vector and atomic operations,
LLVM memory intrinsics such as \texttt{memcpy}, \texttt{memmove},
and \texttt{memset}, and supported libc memory accesses such as \texttt{strdup}
and \texttt{wmemcpy}. Unsupported
intrinsics, inline assembly, and memory effects hidden inside
unsupported uninstrumented calls are coverage boundaries
(Section~\ref{sec:security:coverage}). All optimizations preserve
the security of the original unoptimized transformation.

\subsection{Spatial Safety}
\label{sec:guarantees:why}
\label{sec:security:spatial}

The spatial guarantee follows by composing the invariants.  PTSan stores
object authority in the pointer value itself: an allocation source writes
the object's ID into the high bits, and LLVM pointer-preserving
operations (bitcasts, PHI nodes, selects, loads and stores of pointer
values, and pointer arithmetic that does not overflow into the ID bits)
carry that ID along with the low address, with no separate shadow state.
So if a pointer value is derived from object \texttt{obj} by in-scope
pointer operations, the ID extracted from it remains \texttt{id(obj)}.
Allocation instrumentation establishes the metadata and live-ID
uniqueness invariants, and dereference mediation ensures the memory
operation cannot execute until the check (the interval-containment check
of Section~\ref{sec:design:checks}) has loaded the metadata for that ID
and proved that the full access range lies inside the recorded bounds.
Therefore, under the scope assumptions, a checked access through a
protected tagged pointer cannot reach outside the allocation object from
which the pointer was derived.

For the P1 and P2 primitives, this guarantee is attacker-independent.
The identifier is fixed by the source allocation and travels with the
pointer, so no choice of length, index, or offset escapes the source
object.  A3 gains nothing from understanding the program, A2 gains
nothing from also understanding PTSan, and A1's control over allocation
does not help either: spatial enforcement never depends on which
identifier an object receives, only on the binding between the pointer
and its object's recorded bounds.  Against all three levels, P1 and P2
reduce to accesses within the source object.  The one escape route is a
P2 offset large enough to carry past bit~47 into the identifier field;
that carry changes the identifier, which makes the access a forgery
attempt, covered by the case analysis of the next subsection.

The P1/P2 split is where location-based checking ends: redzones stop
a P1 sweep at the object boundary, but a P2 access that jumps into
another live object passes an address-validity check, and corrupting
non-control data reachable this way suffices for arbitrary
computation~\cite{hu2016dop}.  PTSan does not distinguish the two
primitives: both reduce to the same interval-containment check
against the source object's bounds.

\subsection{Pointer Forgery and Address-Space}
\label{sec:security:forgery}

The P3 primitive removes the dataflow assumption: if the attacker can
write the high bits directly, a pointer's identifier no longer
necessarily came from its source allocation, but the runtime check is
still executed.  A forged high-bit identifier access of \texttt{addr}
has four possible outcomes.  An unused identifier fails because its
bounds are empty; a live identifier fails if \texttt{addr} is outside
of the object it names; the zero and default identifiers fail in
strict mode; and the only passing case is when the identifier belongs
to an object that contains \texttt{addr}.  In that last case the
attacker has reconstructed a valid pointer to that object, not an
arbitrary-address write primitive.

A passing forgery therefore requires a correct identifier--address
pair, and the attacker levels differ in how they can obtain one.  A3
does not know this is a possibility and thus is blocked.  A2 knows
PTSan's layout but cannot predict identifier assignment, and under
ASLR it does not know where any given object lives; recovering either
piece of information requires an information leak, which on these
programs typically means an out-of-bounds read, exactly the access
class PTSan already blocks.  A1 can do better: by controlling all
allocation it can predict which identifier names which object, so it
can supply a valid identifier.  But it still needs the object's
address, so for heap objects A1 stands where A2 stands: blocked until
ASLR is defeated.  This is a stronger position than ASLR alone
provides.  Under ASLR without pointer authority, an attacker who can
write arbitrary addresses can still groom the heap and corrupt
whatever lands at a chosen address; under PTSan, a write succeeds only
if the attacker already knows the specific object and address it
targets.

Stack objects are the exception.  ASLR randomizes the stack base but
not the relative layout of stack slots, so attackers A1 and A2, who
can forge a stack identifier and compute addresses relative to a stack
pointer they hold, may reach another stack variable in the same region.
PTSan still confines such an attacker to user-declared stack
variables, which excludes saved return addresses, so it does not yield
direct control-flow hijacking.  Combining PTSan with stack-layout
randomization~\cite{aga2019smokestack,lee2022savior} would recover for
stack objects the property ASLR gives heap objects.

\subsection{Temporal Safety}
\label{sec:guarantees:temporal}
\label{sec:security:temporal}

The P4 primitive exercises the same check but depends on identifier
reuse, and PTSan's temporal protection is probabilistic even under the
full assumption list.  When an object is freed, PTSan checks that the
freed pointer names the object's recorded base address, clears the
bounds entry, and returns the identifier to the runtime pool.  A stale
pointer keeps its old identifier, but the cleared bounds give it a zero
span, so any nonzero-size dereference fails until the identifier is
reused.  A double or invalid free fails for the same reason: the second
deallocation no longer finds a live object with a matching base for
that identifier.

The residual risk is identifier reuse, and the attacker levels differ
exactly in their ability to arrange it. Identifier assignment is
deterministic: the non-stack ID pool is a FIFO ring, so a freed
identifier is not reassigned until the ring cycles, on the order of
$2^{16}$ intervening allocations. Reuse is therefore never immediate,
and short-lived stale pointers, the common case in practice, are
caught with certainty. The same determinism is the worst-case
weakness: A1, which can model every allocation and free, knows which
identifier each allocation receives and can time a victim allocation
to receive the stale identifier at an overlapping address. Learning
the controlled pointer's ID requires either being the only source of
allocation activity for the program's lifetime or an information
leak, which PTSan blocks; this is what makes A1 unrealistic. A2 and A3 cannot
observe the ring's position, so to them the identifier carried by any
object overlapping their stale address is effectively uniform over the
identifier space: hitting the stale identifier requires on the order
of $2^{15}$ to $2^{16}$ allocation-and-probe events, and every failed
probe is a detected violation.

We do not count address overlap as a further line of defense:
re-tenanting the stale address is the premise of exploiting a
use-after-free, and allocators make it routine.  PTSan's barrier is
orthogonal to placement: the re-tenanted victim carries whatever
identifier the pool dispenses while the stale pointer keeps its old
one, so the identifier analysis above is the whole story.

The practical guarantee
is therefore high-probability detection of use-after-free,
double-free, and invalid-free errors against A2 and A3, defeated only
by the highly unrealistic A1.  Randomized identifier reuse would make
the assignment schedule unpredictable even to an attacker who models
all allocation activity; we leave it to future work.

\subsection{Boundaries of the Guarantee}
\label{sec:guarantees:out-of-model}
\label{sec:security:coverage}
\label{sec:security:threads}
\label{sec:security:limitations}

The analysis above covers checked accesses in strict mode.  Four
boundaries delimit it.

\textbf{Optimized checks.}  The proof obligation is for base
instrumentation.  Section~\ref{sec:optimizations} treats every
optimization as a local replacement governed by a preservation rule: a
check may be elided, merged, or hoisted only when LLVM analyses prove
that the replacement covers the same accesses with compatible metadata,
and PTSan falls back to the base check otherwise.  Imprecision in those
analyses therefore costs performance, not safety.  ID-strip hoisting
(Section~\ref{sec:optimizations:strip-hoisting}) relocates tag strips
and retags rather than checks; it is check-neutral because a retag
reattaches the identifier recovered from the pointer's own derivation,
so every check still consults the object named by the original ID.

\textbf{Coverage.}  Where an allocation site is not intercepted, its
objects are untagged (and fail closed in strict mode); where a memory
access is not instrumented (the coverage boundaries of
Section~\ref{sec:guarantees:invariants}), it is unchecked.
Protection degrades only for the specific untagged pointer or
unchecked access, and many libc routines are covered by wrappers that
validate pointer arguments before handing them to uninstrumented code.
Identifier exhaustion is bounded the same way: when more than $2^{16}$
objects ($2^{15}$ under Linear Address Masking) are live at once,
strict mode fails closed on the next checked access through the
fallback identifier, preserving safety at the cost of stopping the
program.  Section~\ref{sec:evaluation:memory} reports identifier
pressure empirically.

\textbf{Concurrency.}  Spatial checks remain sound in
multithreaded programs: each checked access validates its identifier
and access range against the metadata it observes.  Temporal checks are
weaker under unsynchronized free/use races, because the bounds load and
the check are not one atomic step with the deallocation, and the bounds
load may be hoisted away from the access
(Section~\ref{sec:optimizations}).  A racy use-after-free is a data
race, a class of error outside PTSan's scope, and we treat it as
outside the temporal guarantee. Non-racy temporal errors are unaffected
by multithreaded programs.

\textbf{Granularity.}  PTSan tracks allocation objects, not subobjects,
so intra-object field overflows are outside the spatial guarantee
(Section~\ref{sec:guarantees:model}).  The evaluation measures this
boundary directly (Section~\ref{sec:evaluation:mset}).

\section{Optimizable Memory Safety}
\label{sec:optimizations}

PTSan's base operations are cheap by construction: ID extraction is
a single shift of a register value, the bounds lookup is a direct
index into a flat table, and stack IDs come from thread-local state
(Sections~\ref{sec:design} and~\ref{sec:implementation}).  This
section is about the second half of the overhead story: because
identity propagation is free and bounds loads are separable from the
checks they feed, every remaining check component (ID extraction,
bounds load, range computation, and branch) is ordinary LLVM IR that
the compiler can optimize independently.  The same holds
for the software strips that produce canonical addresses
(Section~\ref{sec:design:pointer-representation}): they too are
ordinary IR, and PTSan places them with a global cost model
(Section~\ref{sec:optimizations:strip-hoisting}).

This section describes those transformations, including the
two-stage pipeline that optimizes loop checks, assuming the
correctness of the LLVM analyses on which PTSan builds: dominance,
postdominance, loop information, target library information, and
Scalar Evolution.  When an analysis cannot prove the
condition an optimization requires, PTSan falls back to a less
optimized check, so imprecision affects performance, not the base
guarantee (Section~\ref{sec:base-correctness}).

\subsection{Analysis Framework}
\label{sec:optimizations:analysis}

PTSan first reduces each memory access to four facts: the base pointer
from which the access is derived, the location where bounds may be
loaded, the range of bytes that the access may touch, and the set of
program points where a check for that access could be inserted.

\textbf{Base-pointer analysis.}  For each access operand, PTSan
traces the LLVM value from which the operand is derived, the access's
\emph{base pointer}, using LLVM's \texttt{getUnderlyingObject} plus
handling for PHI nodes and selects: if all incoming values share a
base, that base is used; otherwise the PHI/select itself is.  The
analysis is best effort, since the access pointer always remains a
valid base.

\textbf{Range analysis.}  PTSan represents each memory operation as a
symbolic byte interval: the size comes from the LLVM type or the
intrinsic's length operand, and the lower endpoint is the Scalar
Evolution expression for the access pointer when it can be expanded
safely at a candidate insertion point, or the access pointer itself
otherwise.  For loops, PTSan represents the whole loop as one range,
from the first access through the last access plus the access width,
when Scalar Evolution can compute that extent.  These range objects
are what the pass compares, merges, serializes during loop
preanalysis, and finally materializes at the selected check location.

\textbf{Placement analysis.}  A check insertion point must satisfy
three conditions: it must dominate every access it covers; the bounds
load must dominate it and the range must be expandable there; and the
access must postdominate it, so a hoisted check cannot reject a path
that would not have executed the access.  When hoisting a check out
of a loop, PTSan also requires the access to execute on every
relevant iteration; otherwise a preheader check could reject an
iteration in which the guarded access would not have run.

\subsection{Local Preservation Rule}
\label{sec:optimizations:preservation}

The optimizations below are not proven one at a time.  Instead, they are
all instances of a single local preservation rule, stated here so the
per-transformation descriptions can simply point to the clause they rely
on.  Replacing a set of base checks with a new check is valid when all of
the following hold:

\begin{enumerate}
  \item The replacement check uses the same object ID as every check it
  covers.

  \item The replacement range is the union of the covered access
  ranges.

  \item The replacement check sits at a legal insertion point, in the
  sense of the placement analysis above, for the covered accesses.
  With one refinement: a \emph{set} of covered accesses whose ranges
  union to the replacement range must postdominate the check.  On any
  path where the check executes, those accesses also execute, so the
  original program would have checked at least the replacement range;
  a conditional access already covered by a guaranteed access need not
  postdominate the check on its own.

  \item The replacement check is dominated by its bounds load, and no
  program point that may change the identifier's metadata (a
  free, resize, or unmap) lies between the bounds load and any
  covered access.
\end{enumerate}

The remaining subsections instantiate this rule.  ID-extraction
hoisting moves
only a pure extraction (the ID is the upper 16 bits of the SSA value)
from each access to the dominating base pointer: in a loop, the low
address of \texttt{buf[i]} changes every iteration but its ID does
not, so repeated shifts and masks leave the loop body.  Bounds-load
hoisting enforces clause 4 by stopping at potentially freeing calls;
check elision is the degenerate case where the replacement is no
check; and check merging and loop hoisting replace several access
ranges with one statically proven range covering their union, sharing
a single bounds load so that clause 4 holds once for every covered
access.  One optimization sits outside the rule: ID-strip hoisting
(Section~\ref{sec:optimizations:strip-hoisting}) moves the
compatibility strips and retags of
Section~\ref{sec:design:pointer-representation} rather than checks,
and its obligation, that every dereference still receives a canonical
address and every check the identifier of the same source object, is
stated there.

\subsection{Hoisting Bounds Loads}
\label{sec:optimizations:bounds-hoisting}

Bounds loads are more constrained than ID extraction.  IDs are part of
SSA values and do not change, but bounds metadata can change when an
object is freed, resized, or unmapped.  PTSan therefore hoists a bounds
load only to a point where the metadata cannot be invalidated on any
path to the checked access, as required by clause 4 of the preservation
rule.

For unfreeable objects, such as stack allocations within their function
lifetime and globals, PTSan can load bounds near the base pointer
definition.  For heap and mmap objects, the pass treats unknown calls as
potential lifetime changes.  It may hoist across LLVM intrinsics and
calls that LLVM marks \texttt{nofree}, and it treats nonreturning calls
as not reaching the checked access.  Otherwise, the nearest potentially
freeing call blocks hoisting.  In control-flow regions with multiple
paths, PTSan searches for the highest insertion point that dominates the
access and from which no blocking call is reachable before the access.

This rule is about temporal precision: spatially, loading an older
bound for the same object cannot let an access escape that object's
allocation interval, but a load hoisted across a free could miss an
intervening lifetime end.  The rule preserves the base temporal
property in the absence of data races; racy free/use executions are
outside the temporal guarantee
(Section~\ref{sec:guarantees:out-of-model}).

For stack and global objects, PTSan eliminates the bounds load
entirely: when the base pointer is an \texttt{alloca} or a global,
the base and size are known statically, so the pass folds both into
the check as constants and the ID extraction and metadata loads
disappear, leaving only the interval comparison.  This is the limit
case of clause 4, since the metadata for these objects cannot change
within their lifetime.

\subsection{Conservative Check Elision}
\label{sec:optimizations:elision}

PTSan eliminates a check only when enforcement cannot need it.
Zero-size checks are removed, since the runtime treats them as
vacuously safe.  Checks are also removed for accesses that LLVM
proves dereferenceable and aligned, but only when the base object is
unfreeable (stack objects, constants, and supported global-derived
pointers): eliding a provably in-bounds \emph{heap} check could
remove the detection of a use-after-free before ID reuse, so the
unfreeable condition keeps elision correctness-preserving rather than
a spatial-only shortcut.

\subsection{Merging Checks}
\label{sec:optimizations:merging}

Many programs perform several nearby accesses through the same object
ID.  PTSan merges such checks when one replacement range can cover all
of the original accesses at a legal insertion point.

\textbf{Subsuming merge.}  If two checks use the same base pointer and
the same bounds load, and one access range contains the other, the
larger range can cover both accesses.  The merged check is placed at a
point that dominates the covered accesses and is postdominated by them
according to the placement analysis above.  This case is common in
load/store pairs.

\textbf{Orderable merge.}  If PTSan can prove that two ranges are
ordered, it may replace them with one interval from the lower range
start to the upper range end.  This covers unrolled sequences such as
\texttt{buf[i]} through \texttt{buf[i+3]} with one combined range.
The comparison and merge happen at compile time using Scalar
Evolution; PTSan never adds runtime tests to decide whether checks
can merge.

Both merge strategies require the merged checks to share a bounds load:
checks are grouped by base pointer and bounds-load location, which
establishes clause 4 for the merged check.  No covered access can
observe metadata loaded under different lifetime assumptions, because
every covered access is checked against the same load and no
metadata-changing point separates that load from the access.

\newsavebox{\stripphibasebox}
\begin{lrbox}{\stripphibasebox}
\begin{minipage}[b]{0.37\textwidth}
\begin{lstlisting}[language={},numbers=none,xleftmargin=0.2em,framexleftmargin=0.2em]
f(%p):
 entry:
  br loop
 loop:
  %q = phi [%p, entry], [%next, loop]
  %s = strip %q
  %v = load %s
  %next = gep %q, 1
  br ..., loop, exit
 exit:
  %out = phi [%q, loop]
  call @g(%out)
  ret %out
\end{lstlisting}
\end{minipage}
\end{lrbox}
\newsavebox{\stripphihoistbox}
\begin{lrbox}{\stripphihoistbox}
\begin{minipage}[b]{0.37\textwidth}
\begin{lstlisting}[language={},numbers=none,xleftmargin=0.2em,framexleftmargin=0.2em]
f(%p):
 entry:
  %p.id = get_id %p
  %p.s = strip %p
  br loop
 loop:
  %q.s = phi [%p.s, entry], [%next.s, loop]
  %v = load %q.s
  %next.s = gep %q.s, 1
  br ..., loop, exit
 exit:
  %out.s = phi [%q.s, loop]
  %out.t = tag %out.s, %p.id
  call @g(%out.t)
  ret %out.t
\end{lstlisting}
\end{minipage}
\end{lrbox}

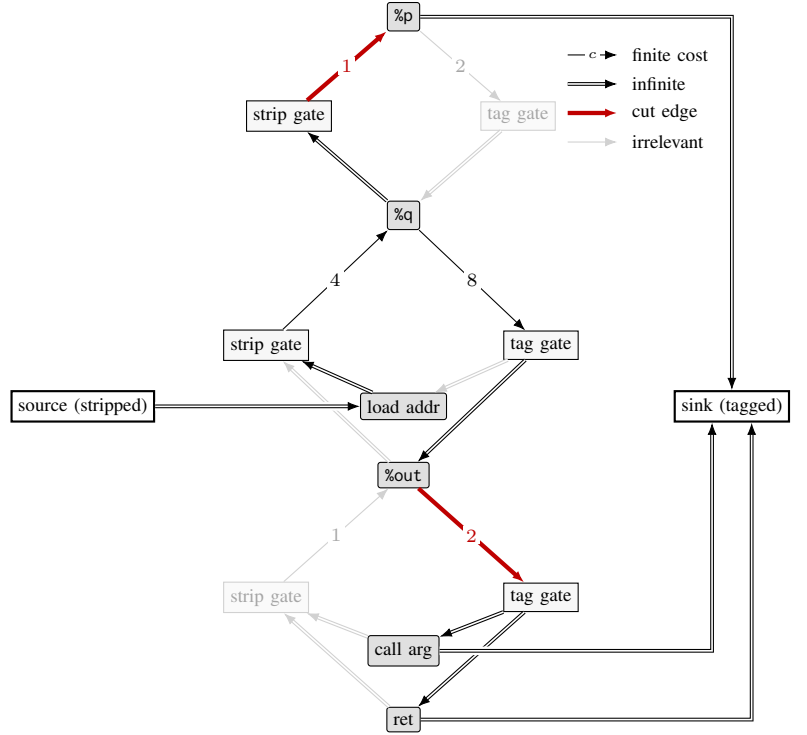
\begin{figure*}[t]
  \centering
  \begin{minipage}[t]{0.39\textwidth}
    \vspace{0pt}
    \centering
    \subfloat[Base placement: strip \texttt{\%q} in the loop.%
      \label{fig:strip-base}]{%
      \usebox{\stripphibasebox}}
    \\[0.8em]
    \subfloat[Hoisted: strip \texttt{\%p}, retag \texttt{\%out}.%
      \label{fig:strip-hoisted}]{%
      \usebox{\stripphihoistbox}}
  \end{minipage}
  \hfill
  \begin{minipage}[t]{0.58\textwidth}
  \vspace{0pt}
  \centering
  \subfloat[Minimum-cut solution; irrelevant edges are faded.%
    \label{fig:strip-cut}]{%
  \resizebox{\linewidth}{!}{%
  \begin{tikzpicture}[font=\scriptsize, >={Latex[length=1.6mm]},
      every node/.style={inner sep=2.5pt},
      edge label/.style={midway, fill=white, inner sep=1pt, font=\tiny},
      cut edge/.style={draw=red!75!black, ultra thick},
      cut label/.style={edge label, text=red!75!black,
        font=\tiny\bfseries},
      finite over/.style={preaction={draw=white, line width=2.2pt}},
      infinite over/.style={preaction={draw=white, line width=3.2pt}},
      irrelevant/.style={draw=black!18},
      irrelevant label/.style={edge label, text=black!35},
      irrelevant node/.style={draw=black!18, fill=black!2,
        text=black!35}]
    \node[draw, fill=black!12, rounded corners=1pt] (p) at (0,6.6)
      {\texttt{\%p}};
    \node[irrelevant node] (ptg) at (1.5,5.3) {tag gate};
    \node[draw, fill=black!3] (psg) at (-1.5,5.3) {strip gate};
    \node[draw, fill=black!12, rounded corners=1pt] (q) at (0,4.0)
      {\texttt{\%q}};
    \node[draw, fill=black!3] (qtg) at (1.8,2.3) {tag gate};
    \node[draw, fill=black!12, rounded corners=1pt] (load) at (0,1.5)
      {load addr};
    \node[draw, fill=black!12, rounded corners=1pt] (out) at (0,0.6)
      {\texttt{\%out}};
    \node[draw, fill=black!3] (qsg) at (-1.8,2.3) {strip gate};
    \node[draw, fill=black!3] (otg) at (1.8,-1.0) {tag gate};
    \node[draw, fill=black!12, rounded corners=1pt] (call) at (0,-1.7)
      {call arg};
    \node[draw, fill=black!12, rounded corners=1pt] (ret) at (0,-2.6)
      {ret};
    \node[irrelevant node] (osg) at (-1.8,-1.0) {strip gate};
    \node[draw, thick] (src) at (-4.2,1.5) {source (stripped)};
    \node[draw, thick] (sink) at (4.3,1.5) {sink (tagged)};
    \path (sink.south west) -- coordinate[pos=0.33] (sinkcall)
      (sink.south east);
    \path (sink.south west) -- coordinate[pos=0.67] (sinkret)
      (sink.south east);
    \begin{scope}[shift={(2.15,6.1)}]
      \draw[->] (0,0) -- node[edge label] {$c$} (0.65,0);
      \node[anchor=west] at (0.75,0) {finite cost};
      \draw[->, double] (0,-0.38) -- (0.65,-0.38);
      \node[anchor=west] at (0.75,-0.38) {infinite};
      \draw[->, cut edge] (0,-0.76) -- (0.65,-0.76);
      \node[anchor=west] at (0.75,-0.76) {cut edge};
      \draw[->, irrelevant] (0,-1.14) -- (0.65,-1.14);
      \node[anchor=west] at (0.75,-1.14) {irrelevant};
    \end{scope}
    \draw[->, irrelevant] (p) --
      node[irrelevant label, font=\scriptsize] {$2$} (ptg);
    \draw[->, double, irrelevant] (ptg) -- (q);
    \draw[->, double, irrelevant] (qtg) -- (load);
    \draw[->, double, irrelevant] (out) -- (qsg);
    \draw[->, double, irrelevant] (call) -- (osg);
    \draw[->, double, irrelevant] (ret) -- (osg);
    \draw[->, irrelevant] (osg) --
      node[irrelevant label, font=\scriptsize] {$1$} (out);
    \draw[->, double, infinite over] (src) -- (load);
    \draw[->, double] (p.east) -| (sink.north);
    \draw[->, double] (ret.east) -| (sinkret);
    \draw[->, double] (q) -- (psg);
    \draw[->, cut edge] (psg) --
      node[cut label, font=\scriptsize\bfseries] {$1$} (p);
    \draw[->] (q) -- node[edge label, font=\scriptsize] {$8$} (qtg);
    \draw[->, double] (qtg) -- (out);
    \draw[->, double] (load) -- (qsg);
    \draw[->, finite over] (qsg) --
      node[edge label, font=\scriptsize] {$4$} (q);
    \draw[->, cut edge] (out) --
      node[cut label, font=\scriptsize\bfseries] {$2$} (otg);
    \draw[->, double] (otg) -- (call);
    \draw[->, double] (otg) -- (ret);
    \draw[->, double, infinite over] (call.east) -| (sinkcall);
  \end{tikzpicture}}}
  \end{minipage}
  \caption{ID-strip hoisting.  Finite labels are
  capacities: strip and tag cost 1 and 2, respectively, multiplied by
  4 per loop level. The base placement in (a) cuts \texttt{\%q}'s
  strip-cost edge (4).  The minimum cut (red) instead cuts \texttt{\%p}'s
  strip-cost edge (1) and \texttt{\%out}'s tag-cost edge (2), producing
  (b) with one retag shared by both escaping uses.  }
  \Description{Three panels.  (a) PHI-explicit IR strips the loop-carried
  value on every iteration.  (b) The strip moves to the entry block and
  the exit value is retagged once for the call and return.  (c) The
  corresponding min-cut graph highlights the strip and tag edges in red
  and fades edges irrelevant to this solution.}
  \label{fig:strip-hoist}
\end{figure*}

\subsection{Loop Checks}
\label{sec:optimizations:loop-hoisting}
\label{sec:optimizations:preanalysis}

A local sanitizer check in a hot loop adds arithmetic, loads, and a
branch to every iteration.  PTSan uses Scalar Evolution to replace
such checks with one preheader check when it can compute the full
range touched by the loop.

For an access whose address is represented by a Scalar Evolution
add-recurrence, PTSan asks LLVM for the first value, the last value, and
the loop trip count.  If the access moves monotonically upward, the
loop range starts at the first access and extends through the last
access plus the access width.  If it moves monotonically downward, the
endpoints are swapped.  If the step is loop-invariant but the sign is
not statically known, PTSan materializes a loop-range check that selects
the minimum endpoint at runtime and checks the absolute span.  If the
step is not analyzable, the pass leaves the check in the loop.

Hoisting also requires the preheader to be a legal insertion point,
including the every-iteration condition of the placement analysis
(Section~\ref{sec:optimizations:analysis}).  These conditions constrain
only the range \emph{check}: the bounds load it depends on can be
hoisted much more freely, under only the metadata-stability conditions
of Section~\ref{sec:optimizations:bounds-hoisting}.  Decoupling the
bounds load from the check this way is something most sanitizers do
not do.

These hoisting conditions fail on exactly the loops that matter most.
After LLVM vectorizes a loop, its body holds alternative access
sequences, a vector fast path and a scalar remainder, and neither
postdominates the loop body, so most optimized hot loops cannot be
hoisted after the fact.  Instrumenting before vectorization would
restore the simple scalar structure but disrupt the loop optimizations
themselves.  PTSan therefore uses a two-stage pipeline: it analyzes
loop accesses early, while they are still in scalar form, preserves
the results as metadata, and consumes them after LLVM has transformed
the loop.

The early pass does not modify program instructions.  It finds loop
accesses that can be described by base pointer and access range, records
whether an entire loop or an entire base within a loop has been solved,
and serializes the result into loop metadata.  Later, the final PTSan
instrumentation pass consumes that metadata.  If the metadata says that
a loop is completely covered, the pass can ignore the individual memory
operations that remain inside the transformed loop and emit the
precomputed range checks instead.  If only a particular base was solved,
PTSan elides checks for accesses derived from that base and falls back
to local analysis for the rest.

The implementation also preserves PTSan loop metadata across LLVM loop
transforms that clone or restructure loops, including vectorization,
unrolling, runtime unrolling, unroll-and-jam, and loop distribution.
This metadata propagation is an engineering detail, but it is central to
the performance result: PTSan can let LLVM optimize the program first
and still recover the loop-level memory ranges needed for low-overhead
checking.

\subsection{Hoisting ID Strips}
\label{sec:optimizations:strip-hoisting}

The optimizations above reduce the cost of checking; ID-strip hoisting
reduces the cost of compatibility.  Without LAM, every dereference
requires a canonical address
(Section~\ref{sec:design:pointer-representation}), so base
instrumentation strips the identifier with one bitwise mask, immediately
before each memory operation.  Each strip is cheap, but a strip inside
a hot loop executes on every iteration; unhoisted stripping accounts
for roughly 20 percentage points of PTSan's SPEC
CPU 2017 overhead (Section~\ref{sec:evaluation:ablation}).

\begin{figure}[t]
  \centering
  \footnotesize
  \begin{minipage}{0.96\columnwidth}
  \begin{algorithmic}
    \Procedure{BuildGraph}{$F$}
      \State $G \gets$ graph containing each pointer value in $F$
      \State $S_{\mathrm{strip}},T_{\mathrm{tag}} \gets
        \Call{NewNode}{G},\Call{NewNode}{G}$
      \ForAll{pointer-valued SSA definitions $v$ in $F$}
        \If{$v$ originates a pointer}
          \State \Call{Edge}{$v,T_{\mathrm{tag}},\infty$}
          \State \textbf{continue}
        \EndIf
        \If{$v$ is a PHI or select over pointer inputs $X$}
          \ForAll{$x \in X$}
            \State \Call{EnsureGates}{$G,x$}
            \State \Call{Edge}{$T(x),v,\infty$}
            \State \Call{Edge}{$v,S(x),\infty$}
          \EndFor
          \State \textbf{continue}
        \EndIf
        \State $x \gets$ pointer operand of $v$
        \State \Call{EnsureGates}{$G,x$}
        \If{$v$ is \Call{Tagged}{$x$}}
          \State \Call{Edge}{$v,T_{\mathrm{tag}},\infty$}
          \State \Call{Edge}{$T(x),v,\infty$}
          \State \Call{Edge}{$v,S(x),\infty$}
        \ElsIf{$v$ is \Call{Strip}{$x$}}
          \State \Call{Edge}{$S_{\mathrm{strip}},v,\infty$}
          \State \Call{Edge}{$S_{\mathrm{strip}},S(x),\infty$}
        \ElsIf{$v$ is transparent with pointer operand $x$}
          \State \Call{Edge}{$T(x),v,\infty$}
          \State \Call{Edge}{$v,S(x),\infty$}
        \EndIf
      \EndFor
      \State \Return $G$
    \EndProcedure
    \Procedure{EnsureGates}{$G,v$}
      \If{$v$ has no conversion gates}
        \State $T(v),S(v) \gets \Call{NewNode}{G},\Call{NewNode}{G}$
        \State \Call{Edge}{$v,T(v),\Call{TagCost}{v}$}
        \State \Call{Edge}{$S(v),v,\Call{StripCost}{v}$}
      \EndIf
    \EndProcedure
  \end{algorithmic}
  \end{minipage}
  \caption{Construction of the polarity min-cut graph.  Pointer SSA
  values are graph nodes, and $T(v)$ and $S(v)$ are value $v$'s shared
  tag and strip conversion gates.  Hard consumers have already been
  made explicit as \textsc{Tagged} and \textsc{Strip} SSA anchors.
  Conversion costs include loop weighting.}
  \label{fig:strip-graph-construction}
\end{figure}

Figure~\ref{fig:strip-hoist} shows ID-strip hoisting before and after. In
Figure~\ref{fig:strip-base}, the original base placement strips the
loop-varying address \texttt{\%q} on every iteration.  In
Figure~\ref{fig:strip-hoisted}, the strip has been hoisted
through the address computation to the loop-invariant base
\texttt{\%p} and executes once, in the entry block; the address the
loop loads from, \texttt{\%q.s} derived from \texttt{\%p.s}, is
already canonical.  The price
appears after the loop: \texttt{\%out} is now stripped but escapes at
the call and the return, where PTSan requires tagged values, so a
\emph{retag} must re-attach the identifier saved at entry.  Here one
retag serves both escaping uses, and the exchange is a clear win once
the loop runs more than a few iterations.  Blind ID-strip hoisting is not
necessarily a win in
all circumstances: a retag forced \emph{inside} a loop can cost more than the
strips it saves.  Strip placement is therefore a global cost problem
rather than a peephole rule, and the ablation bears this out: on SPEC,
a local placement rule recovers less than half of what the global
formulation below does (Section~\ref{sec:evaluation:ablation}).

PTSan formulates the problem over polarities.  Each pointer-typed SSA
value is assigned a \emph{polarity}, stripped or tagged, subject to
two constraint sets.  Uses are constrained individually: dereference
address operands need stripped values, while identifier extractions
feeding checks, pointer-to-integer casts, and each point where a
pointer escapes local dataflow (return values, call arguments, and
pointer values stored to memory) need tagged values.  Values are
constrained by recoverability: only pass-through definitions
(\texttt{getelementptr} address computations, PHI nodes, selects, and
similar) may lose their tagged version.  This is because a retag can
recover their identifier from a dominating tagged ancestor, as the
retag of Figure~\ref{fig:strip-hoisted} recovers \texttt{\%out}'s
identifier from \texttt{\%p}. For a PHI or select, PTSan synthesizes a
PHI or select over the ancestors' identifiers.  Values that originate
pointers, such as arguments, call results, and loaded pointer values,
stay tagged: stripping them would discard the only copy of the
identifier.

Choosing polarities is then a two-terminal minimum cut: a source
representing stripped, a sink representing tagged, a node per SSA
value, and the constraints above as infinite terminal edges
(need-stripped uses tie to the source; need-tagged uses and
pointer-originating values tie to the sink).  The subtlety is how to charge for
conversions.  Connecting each value directly to each of its users
would charge per boundary-crossing use: a stripped \texttt{\%out} in
Figure~\ref{fig:strip-base} would pay for two retags,
although the rewrite materializes one.  PTSan instead routes every
use through two per-value \emph{gate} nodes representing the option
of materializing one tagged or one stripped version of the value; in
Figure~\ref{fig:strip-cut}, each value owns such a pair.  The
use edges are infinite-cost, so a cut can separate a value from
differently-polarized uses only by crossing a finite gate edge, and
however many uses share that conversion, the cut pays its cost once:
the minimum cut prices exactly the conversions the rewrite will
insert.  Figure~\ref{fig:strip-graph-construction} summarizes this
construction.  Figure~\ref{fig:strip-cut} assembles the whole instance for
the example function and shows its minimum cut: the terminal ties
force the load's address use to the source side and the call use,
the return use, and the argument \texttt{\%p} to the sink side, and
the cut crosses exactly two finite edges, the strip of \texttt{\%p}
and the retag of \texttt{\%out}.
(Only edges from the source side to the sink side count in a
directed cut, so the infinite edges that cross back are free.)
Edge weights grow exponentially with loop depth, so
conversions migrate out of loop nests, and retags weigh more than
strips because a retag also synthesizes its identifier and holds it
live, adding register pressure.  PTSan computes the cut with Dinic's
max-flow algorithm~\cite{dinic1970maxflow}, reads each value's polarity off its side of the
cut, inserts the strips and retags the cut edges denote, and removes
conversions the assignment makes redundant.

The pass runs late, after check ranges have been materialized as
ordinary IR, so every producer and consumer of tagged values is
visible to the graph.  It is guarantee-neutral by construction: check
operands are need-tagged uses and pointer-originating values keep
every identifier recoverable, so hoisting changes where conversions
execute, never which object a check consults.  Under LAM, dereferences leave the
need-stripped set, and the same analysis operates on the few strips
that remain at canonical-address boundaries such as inline assembly
and kernel-bound pointers.

\section{Implementation}
\label{sec:implementation}

PTSan is implemented as an LLVM 19 sanitizer: an application is
rebuilt with \texttt{-fsanitize=ptsan}, with no source changes; the
flag runs PTSan's passes and links the static compiler-rt runtime.  PTSan
targets x86-64 and AArch64: the passes operate on LLVM IR and are
architecture-independent.

Two passes implement the pipeline of Section~\ref{sec:optimizations}: the
loop preanalysis pass runs at the vectorizer-start extension point and
writes only metadata, and the main instrumentation pass allocates IDs,
inserts checks, strips IDs where pointers escape to code expecting raw
addresses (placing strips and retags with the min-cut analysis of
Section~\ref{sec:optimizations:strip-hoisting}), and releases stack
IDs on function exit.  The runtime's
metadata tables are parameterized by the ID width (16 bits by default;
the LAM build uses 15 and enables LAM\_U48 at startup) and are backed
by roughly 950 libc interceptors built on LLVM's sanitizer
interception infrastructure: allocation and lifetime wrappers cover
the \texttt{malloc} family, \texttt{mmap}, and \texttt{munmap};
generic wrappers strip pointer arguments and retag returned pointers
that point into an argument; and string, memory,
\texttt{pthread\_create}, and formatting functions have hand-written
support.  The remaining compatibility limits are discussed in
Section~\ref{sec:security:limitations}.  The implementation is about
13{,}700 lines of code across the passes and runtime, plus 7{,}700
lines of tests.  PTSan, comprising the modified LLVM toolchain and
the compiler-rt runtime, is available as open source at
\url{https://github.com/Aarno-Labs/PointerTagSanitizer}.  We intend
to propose PTSan for upstream inclusion in
LLVM: following the structure of sanitizers (instrumentation passes, a
compiler-rt runtime, and the shared interceptor infrastructure) is a
deliberate step toward that goal.

\section{Evaluation}
\label{sec:evaluation}

This section presents results that test whether PTSan
delivers pointer-based memory safety at a deployable cost.  Our
results support four claims:

\begin{enumerate}

\item On stock x86-64 hardware and kernels,
PTSan's runtime overhead is \SpecOvNolamAll{} geomean on SPEC CPU 2017
and \LlvmOvNolamGeo{} on LLVM MultiSource, at parity with the fastest
published redzone sanitizer and roughly a third of the published
overhead of prior pointer-based systems, and hardware address masking
reduces these to \SpecOvLamAll{} and \LlvmOvLamGeo{}.

\item The cost is stable across architectures: the same suites measure
\SpecOvArmAll{} and \LlvmOvArmGeo{} on ARM64 without hardware address masking.

\item PTSan's physical memory overhead is \MemPtsanAll{} geomean, effectively native, where
ASan and RSan cost \MemAsanAll{} and \MemRsanAll{}.

\item On MSET, PTSan matches the originally published detection
coverage of SoftBound+CETS, the strongest prior pointer-based system,
including categories that no location-based sanitizer detects; only
SoftBound+CETS's revisited port does better, and only on the intra-object
cases which PTSan does not target.

\end{enumerate}

\subsection{Experimental Setup}
\label{sec:evaluation:setup}

\textbf{Hardware and toolchain}
x86-64 runtime-overhead experiments (SPEC, LLVM MultiSource, the
ablation, and the server workloads) run on an Intel Core Ultra~9 285H
machine with 96\,GB of memory and Ubuntu 24.04.  ARM64 experiments run
on an AWS m7g.2xlarge instance: an AWS Graviton3 (Arm Neoverse-V1,
eight physical cores, no SMT, fixed 2.6\,GHz clock) with 32\,GiB of
memory and Ubuntu 24.04, with each benchmark pinned to a single core
and reported as the median of three runs.  All native and instrumented
builds use \texttt{-O3}.  The memory overhead, RSan, and MSET
experiments run on a separate AMD Ryzen~9 9950X3D machine with 60\,GiB
of memory and Ubuntu 24.04, with each benchmark pinned to a fixed
physical core under the \texttt{performance} governor.

\textbf{Configurations}
We report three configurations, named by architecture.  The default,
\emph{PTSan-x86}, enables all the optimizations of
Section~\ref{sec:optimizations}, strips identifier bits in software
where uninstrumented code requires canonical addresses, and supports
$2^{16}$ live object IDs; it runs on stock hardware and
kernels.

\emph{PTSan-LAM} instead uses Intel Linear Address Masking
to elide most stripping, with a $2^{15}$ ID budget. We report
PTSan-LAM as a secondary configuration rather than the headline
because of its deployment caveats.  PTSan needs the LAM\_U48
variant~\cite{intel-lam}, which is not in mainline Linux: the
mainlined LAM\_U57 exposes too few maskable bits for PTSan's ID
budget, and address masking has drawn security concerns since SLAM
showed it enlarges the Spectre gadget surface~\cite{hertogh2024slam}.
Our measurements use a custom Linux 6.1 kernel.  As address-masking
hardware and kernel support mature, the PTSan-LAM numbers indicate
where PTSan's overhead is headed.

\emph{PTSan-ARM} is the PTSan-x86 configuration compiled for AArch64, where
stripping is always in software.  Where the architecture does not
matter, we write simply PTSan.  All runtime-overhead measurements use
PTSan's default stack-ID handling: the thread-local fast path reuses
stack identifiers immediately and does not clear their bounds entries
at function exit.

We compare against AddressSanitizer (ASan)~\cite{serebryany2012asan}
and RangeSanitizer (RSan)~\cite{gorter2025rangesanitizer} measured
locally, and against the published overheads of
SoftBound+CETS~\cite{nagarakatte2009softbound,nagarakatte2010cets} and
CUP~\cite{burow2018cup}.  Following its upstream methodology, RSan is
measured against its own \texttt{-O2} baseline built with the same
RSan LLVM build; its runs cover SPEC's pure-C/C++ rate benchmarks
(excluding the mixed-language \texttt{511.povray\_r},
\texttt{526.blender\_r}, and \texttt{507.cactuBSSN\_r}) minus
\texttt{502.gcc\_r}, whose RSan build does not complete,
leaving a 13-benchmark subset.  Each system is therefore reported as
overhead over its own baseline, which removes compiler and machine
differences from the comparison.

\textbf{Benchmarks}
Our CPU evaluation uses two suites.  First, we run SPEC CPU 2017
v1.1.9 \texttt{intrate} and \texttt{fprate} with reference inputs,
base tuning, and one copy; we evaluate the
\SpecBenchCount{} C/C++ benchmarks and exclude benchmarks that execute Fortran
code.  Second, we run the \LlvmTestCount{} LLVM test-suite MultiSource
benchmarks, which the LLVM documentation describes as ``entire
programs with multiple source files'' where ``large benchmarks and
whole applications go''~\cite{llvm-testsuite-guide}.  We exclude
MultiSource programs with native runtime under 10\,ms, which are too
noise-sensitive to measure, and five programs whose runs PTSan flags:
in four (\texttt{gs}, \texttt{netbench-url}, \texttt{hbd}, and
\texttt{PENNANT}) PTSan detects real memory-safety
bugs\footnote{We root-caused these violations to assess disclosure
obligations and found no undisclosed vulnerability in maintained
software: the \texttt{gs} and \texttt{PENNANT} bugs are already fixed
in the maintained upstreams (Ghostscript and LANL PENNANT),
and \texttt{netbench-url} and \texttt{hbd} exist only in the public
benchmark corpus, with no maintained upstream.}, and \texttt{siod}
walks the stack across frames by address arithmetic, which violates
PTSan's pointer-authority model (Section~\ref{sec:guarantees:model})
without being a memory-safety bug.  Because flagged runs take PTSan's
intentionally slow report path, they would skew the overhead numbers;
\LlvmMeasuredLam{} programs remain
on x86-64 and \LlvmMeasuredArm{} on ARM64, where
\texttt{security-sha} also falls under the runtime cutoff.
Overheads are reported as the geometric mean of per-benchmark
instrumented/native runtime ratios, always over the same benchmark set
within a comparison.  For memory overhead we report peak physical
memory from cgroup~v2
\texttt{memory.peak}, not virtual reservation, taking the maximum
successful command row per benchmark.  For detection coverage we use
MSET~\cite{vintila2025mset}, described with its detection convention
in Section~\ref{sec:evaluation:mset}.

\subsection{Runtime Overhead}
\label{sec:evaluation:spec}

PTSan provides pointer-based checking at \SpecOvNolamAll{} geomean
overhead on SPEC CPU 2017 on commodity x86-64 systems and
\SpecOvArmAll{} on ARM64, at parity with the fastest redzone
sanitizer.  ID-strip hoisting
(Section~\ref{sec:optimizations:strip-hoisting}) recovers most of the
software tag-stripping cost on stock hardware; hardware address
masking removes what remains, bringing the x86-64 geomean to
\SpecOvLamAll{}.

\subsubsection{SPEC CPU 2017}

\begin{table}[t]
  \centering
  \caption{Runtime overhead and peak live objects on SPEC CPU
  2017 (\dag: exceeds the $2^{16}$ ID budget).}
  \label{tab:spec-runtime}
  \footnotesize
  \resizebox{\columnwidth}{!}{\begin{tabular}{@{}lrrrr@{}}
  \toprule
  Benchmark & PTSan-x86 & PTSan-LAM & PTSan-ARM & Max live IDs \\
  \midrule
  \multicolumn{5}{@{}l}{\emph{SPECrate 2017 Integer}} \\
  500.perlbench\_r & 181.4\% & 117.2\% & 131.1\% & 1,253,094\textsuperscript{\dag} \\
  502.gcc\_r & 109.4\% & 61.4\% & 95.4\% & 1,076,541\textsuperscript{\dag} \\
  505.mcf\_r & 36.2\% & 32.3\% & 49.6\% & 15,010 \\
  520.omnetpp\_r & 73.5\% & 61.8\% & 17.0\% & 2,388,303\textsuperscript{\dag} \\
  523.xalancbmk\_r & 75.8\% & 57.7\% & 76.7\% & 2,338,108\textsuperscript{\dag} \\
  525.x264\_r & 44.8\% & 40.6\% & 75.2\% & 3,498 \\
  531.deepsjeng\_r & 56.9\% & 49.8\% & 72.5\% & 650 \\
  541.leela\_r & 64.2\% & 57.3\% & 125.8\% & 26,002 \\
  557.xz\_r & 34.0\% & 31.9\% & -2.8\% & 42 \\
  \textbf{Geomean (INT)} & \textbf{70.6\%} & \textbf{55.0\%} & \textbf{65.4\%} & \\
  \midrule
  \multicolumn{5}{@{}l}{\emph{SPECrate 2017 Floating Point}} \\
  507.cactuBSSN\_r & 57.2\% & 57.2\% & 35.9\% & 18,655 \\
  508.namd\_r & 73.9\% & 62.9\% & 34.1\% & 3,492 \\
  510.parest\_r & 39.0\% & 39.3\% & 44.4\% & 2,375,107\textsuperscript{\dag} \\
  511.povray\_r & 136.5\% & 118.3\% & 170.6\% & 11,869 \\
  519.lbm\_r & -2.1\% & -3.1\% & -12.8\% & 2 \\
  526.blender\_r & 18.0\% & 8.8\% & 37.8\% & 342,174\textsuperscript{\dag} \\
  538.imagick\_r & 45.4\% & 36.7\% & 54.9\% & 480 \\
  544.nab\_r & 17.9\% & 12.4\% & 36.0\% & 80,183\textsuperscript{\dag} \\
  \textbf{Geomean (FP)} & \textbf{43.3\%} & \textbf{37.3\%} & \textbf{43.5\%} & \\
  \midrule
  \textbf{Geomean (combined)} & \textbf{57.2\%} & \textbf{46.4\%} & \textbf{54.7\%} & \\
  \bottomrule
\end{tabular}
}
  \Description{Table of per-benchmark PTSan runtime overheads on SPEC
  CPU 2017 on x86-64 with and without LAM and on ARM64, plus peak
  live object-ID counts.}
\end{table}

Table~\ref{tab:spec-runtime} reports per-benchmark overhead for all
three configurations.  PTSan-x86's geomean overhead is
\SpecOvNolamInt{} on the integer suite and \SpecOvNolamFp{} on
floating point, for a combined \SpecOvNolamAll{}.  Every benchmark
builds, runs, and validates on both architectures; the worst cases
are \texttt{500.perlbench\_r} at 181\% on x86-64 and
\texttt{511.povray\_r} at 171\% on ARM64.

The outliers have a consistent shape.  For \texttt{500.perlbench\_r},
profiling attributes the cost to inline-check instruction
growth rather than runtime-library time or locality effects:
instrumentation grows Perl's regex interpreter \texttt{S\_regmatch}
about 9$\times$ in code size, that one function dominates PTSan
execution samples while runtime helpers stay under 2\%, and IPC rises.
Large interpreters over pointer-rich state are close to a worst case
for any pointer-based instrumentation.
\texttt{511.povray\_r}'s overhead (136\% on x86-64) instead concentrates in
stack-object ID setup for its many small, hot functions with
address-taken temporaries, consistent with the stack-ID rung of the
ablation (Section~\ref{sec:evaluation:ablation}).

PTSan-LAM isolates the software tag-stripping cost that remains after
ID-strip hoisting: with stripping elided in hardware, the combined
geomean falls from \SpecOvNolamAll{} to \SpecOvLamAll{}.  Of the
roughly 20-point gap between the pipeline without strip hoisting
(\AblNoStripHoistAll{}, Section~\ref{sec:evaluation:ablation}) and
PTSan-LAM, the compiler now recovers about half in software, which
validates the design decision to keep stripping separable and
optimizable (Section~\ref{sec:optimizations}).

PTSan-ARM shows the cost is portable.  The pipeline compiles
unmodified for AArch64 (all PTSan operations are IR-level, so the only
difference is in the LLVM-backend-generated assembly), and its combined geomean is
\SpecOvArmAll{} against PTSan-x86's \SpecOvNolamAll{}.
Per-benchmark costs diverge more
than the aggregates suggest, and while we did not profile the ARM64
runs to attribute them individually, the pattern tracks per-check ISA
cost.  The x86-64 check sequence is compact: the bounds-table access
folds into one complex-addressing load, and the compare and branch
macro-fuse; AArch64 needs a few more discrete instructions per check.
Call-dense, check-dense, recursion-shaped code amplifies that
difference and loop-shaped code hides it:
\texttt{541.leela\_r} (recursive tree search) and
\texttt{511.povray\_r} (stack-ID-heavy) degrade on ARM64, while
streaming floating-point codes such as \texttt{508.namd\_r} run
cheaper there.

For context, ASan's combined geomean on the same machine is
\AsanOvAll{}, roughly double PTSan-x86's overhead and 2.5$\times$
PTSan-LAM's.  RSan, the fastest published redzone-style sanitizer,
reports 44.5\% on SPEC CPU 2017 speed with its default
implicit-tagging design~\cite{gorter2025rangesanitizer}; our local
runs of that design measure \RsanOvAll{} against its own baseline on
its 13-benchmark feasible subset.  On that subset PTSan-x86's geomean
is \SpecOvNolamCommonRsan{} and PTSan-LAM's is \SpecOvLamCommonRsan{}:
on stock hardware PTSan runs at parity with RSan, and with hardware
masking it is faster.  That comparison gives RSan no hardware
disadvantage: RSan's own address-masking experiments show that explicit
tagging on Intel LAM runs slightly \emph{slower} than its implicit
default (54.0\% vs.\ 51.0\% on SPEC CPU 2006), masking instead
reducing its memory overhead~\cite{gorter2025rangesanitizer}.

Pointer-based checking no longer trades speed for its guarantee:
object-identity detection that no location-based sanitizer provides
(Section~\ref{sec:evaluation:mset}) and 3$\times$ lower memory
footprint (Section~\ref{sec:evaluation:memory}) now come at
redzone-sanitizer speed.  The published overheads of the pointer-based
systems with comparable guarantees, 161\% for SoftBound+CETS and 158\%
for CUP, are roughly triple PTSan's default configuration; those
numbers come from older SPEC suites, so we treat them as indicative
rather than directly comparable.

\subsubsection{LLVM MultiSource}
\label{sec:evaluation:llvm}

\begin{table}[t]
  \centering
  \caption{Overhead on LLVM MultiSource; the ID-budget row counts all
  \LlvmTestCount{} programs.}
  \label{tab:llvm-summary}
  \resizebox{\columnwidth}{!}{\begin{tabular}{@{}lrrr@{}}
  \toprule
   & PTSan-x86 & PTSan-LAM & PTSan-ARM \\
  \midrule
  Programs measured & 149 & 149 & 148 \\
  Geomean overhead & 31.5\% & 27.2\% & 40.8\% \\
  Median overhead & 24.7\% & 21.2\% & 34.9\% \\
  Programs under 50\% overhead & 103 (69\%) & 114 (77\%) & 84 (57\%) \\
  Within ID budget ($2^{16}$ / $2^{15}$ / $2^{16}$) & 152/167 (91\%) & 149/167 (89\%) & 152/167 (91\%) \\
  \bottomrule
\end{tabular}
}
  \Description{Summary statistics of PTSan overhead on LLVM
  MultiSource benchmarks on x86-64 with and without LAM and on
  ARM64.}
\end{table}

The MultiSource suite tests breadth: a large set of C and C++ programs
across many application domains rather than a tuned benchmark kernel
set.  Table~\ref{tab:llvm-summary} summarizes the results; the
distribution appears in Figure~\ref{fig:llvm-cdf}
(Appendix~\ref{app:eval}).  PTSan-x86's geomean overhead is
\LlvmOvNolamGeo{} (\LlvmOvLamGeo{} with LAM, \LlvmOvArmGeo{} on
ARM64).  The median, \LlvmOvNolamMed{}, is below the geomean, and
\LlvmUnderFiftyNolamPct{} of programs stay under 50\% overhead, so
the aggregate is not driven by a few favorable outliers. The tail is
the same pointer-dense shape as the SPEC outliers
(\texttt{kimwitu++}, \texttt{sqlite3}).  On ARM64 the tail shifts, led
by \texttt{lambda} (534\%) and \texttt{sqlite3} (253\%).
\texttt{lambda}, a small lambda-calculus interpreter, is the extreme
of the recursion-shaped pattern above: it executes 3.8 billion checks,
nearly all single-access rather than hoistable loop-range checks, in a
roughly two-second run, a dependent pointer-chasing chain in which
checks are the majority of the dynamic instruction stream, leaving no
slack to hide the added per-check instructions on AArch64.  SPEC
overheads are higher than MultiSource overheads because the SPEC
kernels spend more of their time in long, pointer-dense hot loops; the
MultiSource programs look more like the application code a sanitizer
would actually be deployed on.

\subsubsection{Real-World Servers}
\label{sec:evaluation:nginx}

To test deployment-style workloads, we ran PTSan on seven real-world
server and cryptographic workloads from the Phoronix Test
Suite~\cite{phoronix}, each in its default configuration, on the Intel
machine of Section~\ref{sec:evaluation:setup}
(Table~\ref{tab:servers}; per-workload descriptions in
Appendix~\ref{app:servers}).  Overhead tracks how compute-bound each
workload is: Apache is within measurement noise, and OpenSSL, SQLite,
and memcached stay at or below 9\% with PTSan-x86, while the more
allocation- and request-processing-intensive PostgreSQL, Redis, and
NGINX reach 16\%, 26\%, and 44\%, respectively.  Hardware masking's
benefit is uneven: it nearly eliminates NGINX's overhead (44\% to
2\%), whose cost is dominated by tag stripping, but barely changes
Redis's, leaving a PTSan-LAM maximum of 25\%.  Peak live-ID demand
never exceeds \ServerMaxId{} (Redis), about a quarter of the $2^{16}$
budget, so no workload approaches exhaustion.  These results
demonstrate PTSan's applicability to real-world server workloads at
deployable overhead.

\subsection{Memory Overhead}
\label{sec:evaluation:memory}

PTSan's physical-memory overhead on SPEC is \MemPtsanAll{} geomean
with a maximum of \MemPtsanMax{}: effectively native.  On the same
benchmarks and baseline, ASan costs \MemAsanAll{} and RSan
\MemRsanAll{} (per-benchmark data in Appendix~\ref{app:eval}).
Restricting the comparison to benchmarks within PTSan's $2^{16}$
identifier budget does not change the story: on the \MemIdFitCount{}
in-budget benchmarks with memory data (the 13-benchmark memory
subset lacks \texttt{507.cactuBSSN\_r} and \texttt{511.povray\_r}, so
\MemIdFitCount{} of the \SpecIdFitCount{} in
Table~\ref{tab:spec-runtime} qualify), RSan costs \MemRsanIdFit{}
where PTSan costs \MemPtsanIdFit{}, so the advantage is not an
artifact of workloads whose identifier demand PTSan cannot cover.  As
memory becomes a first-order cost in production fleets, near-native
memory is the difference between a sanitizer that can ship and one
that cannot.

The contrast is sharpest on small-footprint programs:
\texttt{541.leela\_r} has a 31\,MiB native peak, which redzone and
shadow-memory designs inflate to over a gigabyte (34--37$\times$)
while PTSan reaches \MemPtsanMax{}.  Redzone and shadow costs scale with
the number and size of allocations; PTSan has no redzones, no shadow
memory, no quarantine, and no per-pointer metadata in memory.  Its
metadata is a fixed, workload-independent set of object-ID tables:
about 1.2\,MiB at the default $2^{16}$ budget and a few MiB including
the shared sanitizer interceptor state.  The overhead is therefore
bounded by design rather than scaling with allocations.

\subsubsection{ID Pressure}

The flip side of fixed metadata is a finite ID budget, so we measure
each workload's peak demand for simultaneously live IDs.  On SPEC,
\SpecIdFitCount{} of \SpecBenchCount{} benchmarks stay within the
default $2^{16}$ budget (Table~\ref{tab:spec-runtime}); the
\SpecIdExceedCount{} that exceed it (e.g., \texttt{520.omnetpp\_r} at
2.4M live IDs) keep millions of small allocations live concurrently.
SPEC overstates this pressure for at least \texttt{500.perlbench\_r}
and \texttt{502.gcc\_r}, where the SPEC harness disables garbage
collection, so these numbers are not necessarily indicative of
real-world deployments of those programs.  On LLVM MultiSource,
\LlvmIdFitSixteen{} of \LlvmTestCount{} programs
(\LlvmIdFitSixteenPct{}) fit the $2^{16}$ budget and
\LlvmIdFitFifteen{} (\LlvmIdFitFifteenPct{}) fit the $2^{15}$ LAM
budget.  When demand exceeds the budget, PTSan falls back to shared
IDs and protection degrades for the affected objects
(Section~\ref{sec:design:allocation}); strict deployments can
instead fail closed.  These results indicate that the compact-ID
design covers most workloads outright; because peak ID demand is
cheap to measure, a deployment can determine in advance whether a
workload fits the budget rather than discovering exhaustion in
production.

\subsection{Detection Coverage}
\label{sec:evaluation:mset}

\begin{table}[t]
  \centering
  \caption{MSET detection results.  ASan, RSan, and PTSan are local
  runs; the SoftBound+CETS rows are published results, for the
  original system and its revisited LLVM port.}
  \label{tab:mset-results}
  \scriptsize
  \resizebox{\columnwidth}{!}{%
    \begin{tabular}{@{}lcccccc@{}}
      \toprule
      System & Linear & Non-linear & Type conf. & UAF/UAR & Double-free & Misuse-free \\
      \midrule
      ASan & 70/90 & 18/72 & 18/30 & 8/16 & 4/4 & 20/20 \\
      RSan & 58/90 & 18/72 & 16/30 & 10/16 & 4/4 & 0/20 \\
      SoftBound+CETS~\cite{nagarakatte2009softbound,nagarakatte2010cets} & 72/90 & 54/72 & 24/30 & 16/16 & 4/4 & 20/20 \\
      SoftBound+CETS revisited~\cite{10.1145/3642974.3652285} & 84/90 & 66/72 & 28/30 & 16/16 & 4/4 & 20/20 \\
      PTSan & 72/90 & 54/72 & 24/30 & 16/16 & 4/4 & 20/20 \\
      \bottomrule
    \end{tabular}%
  }
  \Description{Table of MSET detection counts by bug family for ASan,
  RSan, the original and revisited SoftBound+CETS, and PTSan.}
\end{table}

On MSET, PTSan detects every inter-object spatial, non-object
spatial, and temporal case\footnote{In order to catch stack temporal
errors, PTSan needs to be run with
\texttt{-ptsan-stack-temporal-safety}, which adds \(\sim\)3\% overhead
on SPEC CPU 2017 (Table~\ref{tab:stack-temporal}).}, matching the
originally published coverage of SoftBound+CETS, the strongest prior
pointer-based system, at a fraction of its overhead.

MSET separates ordinary address-validity failures from
object-identity failures: an access can land inside mapped, currently
allocated memory and still be illegal for the pointer that performed
it~\cite{vintila2025mset}.  Table~\ref{tab:mset-results} reports the
stock bug-family aggregates under the MSET detection convention (a
test counts as detected if the sanitizer reports the violation or
prevents the benchmark's preconditions from being established),
which is also the convention of the published SoftBound+CETS rows.

In the relation-bucket view, PTSan detects 126/126 inter-object and
24/24 non-object spatial cases, and all temporal cases: 16/16
use-after, 4/4 double-free, and 20/20 misuse-of-free, with temporal
detection deterministic until the corresponding ID is reused
(Section~\ref{sec:guarantees:temporal}).  The 0/42 intra-object
result is by design and matches the scope set in
Section~\ref{sec:security:limitations}.  Because PTSan detects every
inter-object and non-object case, all of its spatial misses in
Table~\ref{tab:mset-results} (18 linear, 18 non-linear, and 6
type-confusion, 42 in total) are exactly these 42 intra-object cases.
The only row that exceeds PTSan is the revisited SoftBound+CETS port,
and its entire advantage is 28 of those 42 intra-object cases,
recovered by narrowing bounds at subobject derivations, a mechanism
PTSan does not attempt.

On non-linear out-of-bounds accesses, the category that exercises
redirection into another live object (the same pattern as the
ImageMagick example of Section~\ref{sec:background:cve-example}),
PTSan detects all supported cases (the undetected cases are
intra-object out-of-bounds accesses), while ASan and RSan detect none
(see Table~\ref{tab:mset-supported} in Appendix~\ref{app:eval} for
supported-only data).  This is a semantic gap, not an implementation
gap: a range or address-validity check cannot reject an access whose
target lies inside some other live object; only a pointer-identity
check can.  The temporal categories show a similar separation.  Our
detection evaluation is benchmark-based; an in-the-wild bug-finding
study is future work.

\subsection{Optimization Ablation}
\label{sec:evaluation:ablation}

Table~\ref{tab:ablation-summary} isolates the contributions of the
mechanisms of Section~\ref{sec:optimizations} as a cumulative
waterfall, measured on the Intel machine of
Section~\ref{sec:evaluation:setup}.  The first row is the full
pipeline with LAM (\AblFullLamAll{}); each subsequent row disables
one more mechanism than the row above, down to unoptimized base
instrumentation (\AblNoOptsAll{}), so each $\Delta$ is the marginal
cost of losing that mechanism given that everything above it in the
table is already gone.  Read bottom-up, the compiler optimizations
together cut PTSan-x86's overhead from 140\% to
57\%, a \(\sim\)60\% reduction.

Four effects account for most of the waterfall.  ID-strip hoisting
(Section~\ref{sec:optimizations:strip-hoisting}) is worth
\AblStripHoistPpSpec{} points on stock hardware.  Most of that benefit,
\AblLocalStripPpSpec{} points, comes from global rather than local strip
placement, providing the empirical case for the global min-cut
formulation.  The stack-ID fast path is a runtime mechanism rather than
an optimization pass; its rung (+13.6 points) is consistent with
\texttt{511.povray\_r}'s stack-heavy profile.  Check merging adds 21.1
points.  The final no-optimization rung is the largest (+27.9 points),
as ID extraction can no longer be shared across derived pointers or
hoisted out of loops.
Because the waterfall is cumulative, mechanisms that overlap show
small marginal deltas once their partner is gone: loop check hoisting
adds only half a point after loop preanalysis has already been
removed, although the two together account for several points.
Per-benchmark data appears in Appendix~\ref{app:eval}.

\subsection{Summary}
\label{sec:evaluation:summary}%

PTSan provides pointer-object authority checking with
\SpecOvNolamAll{} runtime overhead on commodity x86-64 systems,
\SpecOvArmAll{} on ARM64, and \SpecOvLamAll{} with LAM,
\MemPtsanAll{} memory overhead, and full
inter-object and temporal MSET coverage, in a recompile-only
deployment model.  PTSan's runtime overhead matches the best
address-validity system on stock hardware while prior identity-based
systems are roughly triple PTSan's runtime overhead percentage. No
published system occupies this point.

\begin{table}[t]
  \centering
  \caption{Optimization ablation on SPEC CPU 2017: a cumulative
  waterfall in which each row disables one more mechanism than the
  row above it; $\Delta$ is the marginal cost in percentage points of
  combined geomean overhead.}
  \label{tab:ablation-summary}
  \resizebox{\columnwidth}{!}{\begin{tabular}{@{}lrrrr@{}}
  \toprule
  Configuration & INT & FP & Combined & $\Delta$ \\
  \midrule
  \textbf{Full pipeline (PTSan-LAM)} & \textbf{55.0\%} & \textbf{37.3\%} & \textbf{46.4\%} &  \\
  $-$ hardware masking (PTSan-x86) & 70.6\% & 43.3\% & 57.2\% & +10.8 \\
  $-$ global strip placement (local only) & 75.2\% & 49.3\% & 62.5\% & +5.3 \\
  $-$ ID-strip hoisting entirely & 80.2\% & 51.7\% & 66.2\% & +3.7 \\
  $-$ stack-ID fast path & 87.8\% & 71.2\% & 79.8\% & +13.6 \\
  $-$ loop preanalysis & 89.3\% & 82.3\% & 86.0\% & +6.2 \\
  $-$ loop check hoisting & 90.0\% & 82.6\% & 86.5\% & +0.5 \\
  $-$ check merging & 104.6\% & 111.1\% & 107.6\% & +21.1 \\
  $-$ check elision & 107.5\% & 111.1\% & 109.2\% & +1.6 \\
  $-$ bounds-load hoisting & 107.3\% & 117.5\% & 112.0\% & +2.8 \\
  $-$ ID-extraction hoisting (no opts.) & 125.2\% & 157.7\% & 139.9\% & +27.9 \\
  \bottomrule
\end{tabular}
}
  \Description{Table of geomean PTSan overhead as optimizations are
  cumulatively disabled, from the full pipeline with LAM down to no
  optimizations.}
\end{table}

\section{Related Work}
\label{sec:related-work}

Surveys of the sanitizer space divide memory-safety checkers into
\emph{location-based} tools, which track which memory is valid, and
\emph{identity-based} tools, which track which object each pointer is
entitled to access, its \emph{intended
referent}~\cite{song2019sok,vintila2025mset}.  Identity-based tools
subdivide by where bounds live: \emph{per-pointer} tracking attaches
bounds to every pointer value, while \emph{per-object} tracking keeps
one bounds record per allocation and recovers it from the
pointer. PTSan is an identity-based, per-object sanitizer that
recovers metadata through a pointer-carried identifier, with temporal
protection by identifier invalidation.  Table~\ref{tab:design-space}
(Appendix~\ref{app:eval}) summarizes guarantees and costs across both
classes.

\subsubsection*{\textbf{Location-Based Sanitizers}}
AddressSanitizer (ASan) popularized the practical sanitizer model:
compile-time instrumentation, a runtime with allocator integration,
shadow metadata, and high compatibility~\cite{serebryany2012asan}.
Its guarantee is adjacency: redzones detect accesses that land in the
poisoned padding next to an object, so the P2 primitive of
Section~\ref{sec:threat-model}, an offset that jumps over the redzone
into another live object, is not flagged.  MSET quantifies the gap:
ASan detects none of the non-linear out-of-bounds cases and misses
half of the use-after-free cases, since its quarantine only delays
reuse~\cite{vintila2025mset}.  The cost profile for ASan is moderate
runtime overhead, but severalfold memory growth from the redzones,
shadow memory, and quarantine~\cite{song2019sok}.  RangeSanitizer
(RSan) is the strongest current sanitizer in this class: it encodes a size
class in unused pointer bits so one metadata load and one comparison
validate an entire access range, sharply cutting runtime
overhead~\cite{gorter2025rangesanitizer}.  RSan derives temporal
detection from the same check, zeroing an object's stored bound at
free so a stale access fails; this catches use-after-free and
double-free, but not invalid-free, because RSan never validates the
freed pointer the way ASan's allocator does (0/20 versus 20/20 in our
MSET runs).  PTSan answers the provenance question ASan and RSan
cannot, detects temporal errors after quarantine-scale reuse windows,
and keeps memory near native, with no redzones, shadow memory, or
address-space quarantine, at runtime cost matching RSan's
(Section~\ref{sec:evaluation:spec}).

\subsubsection*{Per-Pointer Bounds Tracking}
SoftBound attaches disjoint base and bound metadata to each pointer
and narrows bounds at subobject derivations, which makes per-pointer
tracking the only technique with the conceptual potential for complete
spatial safety, including intra-object
overflows~\cite{nagarakatte2009softbound,vintila2025mset}.  CETS adds
a lock-and-key scheme with never-reused 64-bit keys, giving
deterministic temporal detection even after memory
reuse~\cite{nagarakatte2010cets}.  These are the strongest software
guarantees in the design space, and PTSan does not match them: it
tracks allocation-granular bounds, so intra-object overflows are out
of scope (Section~\ref{sec:guarantees:model}), and its 16-bit
identifiers are reused, so temporal detection is probabilistic
(Section~\ref{sec:security:temporal}).

The cost of the stronger guarantee is what PTSan is designed to avoid.
Every pointer copy, call, and store must propagate disjoint metadata
through shadow structures, which keeps even a modern LLVM port of
SoftBound+CETS among the most expensive software point in
Table~\ref{tab:design-space}~\cite{10.1145/3642974.3652285}.
Compatibility remains the other barrier: Song et al.\ report that
SoftBound+CETS raises false alarms on many SPEC benchmarks because
integer-pointer casts and uninstrumented libraries break metadata
propagation~\cite{song2019sok}.

WatchdogLite reaches 29--45\%
overhead for the same model by adding new hardware
instructions~\cite{nagarakatte2014watchdoglite}; PTSan reaches
comparable overhead with, at most, commodity address
masking.  PTSan keeps the
identity-based check but moves the metadata into the pointer value and
a flat table, so propagation costs nothing, no shadow stack or shadow
memory exists to desynchronize, and checks remain optimizable IR on
commodity hardware.

\subsubsection*{Per-Object Tracking with Encoded or Tagged Pointers}
Low-Fat Pointers derive bounds from the pointer value itself by
segregating the heap and stack into size-class regions, at low
runtime cost and near-native
memory~\cite{duck2016lowfat,duck2017stack}.
The encoding is the guarantee's limit: bounds are the allocation
size class, not the object, so overflows into rounding padding pass
undetected~\cite{vintila2025mset}, there is no temporal protection,
and the assumption that out-of-bounds pointers do not escape breaks
real programs~\cite{song2019sok}.  Delta Pointers fold an
overflow check into spare pointer bits at very low cost, but by design
detect only overflows, with no underflow or temporal
coverage~\cite{kroes2018delta}.  PTSan also keeps pointers
machine-sized and pays a higher runtime cost than these systems, but
its tag indexes exact per-object bounds, so padding offers no refuge
and the same identifier supports temporal checking.

Hardware-assisted tagging compares small colors instead of bounds:
HWASan and Arm MTE tag pointers and memory and trap on a
mismatch~\cite{serebryany2018hwasan,arm-mte}, with tags of 4 to 8
bits making detection probabilistic at collision odds as high as one
in sixteen.  The hardware only carries the tag cheaply in the
pointer; HWASan still loads and compares the memory tag in software
at every access, which RSan measures at 132--268\%
overhead~\cite{gorter2025rangesanitizer}.  Only Arm MTE checks the
tag in hardware, and that needs MTE-capable silicon.  PTSan's 16 bits
are an index, not a color: a check compares the access range against
exact bounds, so spatial enforcement is not probabilistic at all, and
only temporal protection degrades with identifier reuse.

CUP is the closest prior design.  Like PTSan, it uses a
pointer-carried identifier to index exact per-object bounds, and it
supports more simultaneously active identifiers~\cite{burow2018cup}.
A CUP enriched pointer replaces the address with a capability
identifier and a 32-bit offset.  When the identifier is reused, the
stale offset is applied to the new object's base, so the pointer no
longer retains the old address needed to distinguish the two
allocations.  Thus the temporal protection is trivially and
deterministically evadable: for a base pointer, \texttt{malloc(A);
  free(A); malloc(B); use(A)} is not caught. PTSan instead preserves
both the canonical address and the identifier; identifier reuse hides
a stale access only when the replacement object's bounds also cover
the old address.  This representation also explains the performance
difference: CUP must reload the base and reconstruct each enriched
address, whereas PTSan's metadata lookup remains a separable check
that LLVM can optimize and Intel LAM can accelerate.  CUP consequently
incurs roughly three times PTSan's runtime overhead.

\subsubsection*{Garbage-Collected Capability C} Fil-C gives each
pointer an invisible capability and uses a concurrent collector so
that \texttt{free} deterministically disables all pointers to the
freed object~\cite{filc,filc-invisicaps,filc-how}.  This is the
strongest temporal story available for C, at multiples of native
runtime~\cite{filc-invisicaps}, and it requires a garbage-collected
runtime, a new ABI, and rebuilding all linked
code~\cite{filc-runtime}.  PTSan occupies the recompile-only point:
weaker temporal protection and an uninstrumented-library boundary, in
exchange for sanitizer-style deployment at a fraction of the cost.

\section{Future Work}
\label{sec:future-work}

PTSan's compact pointer representation establishes a practical point
in the design space, and its optimization pipeline creates a
foundation for exploring richer representations.  We are investigating
larger authority spaces and new identifier-management schemes that
support more simultaneously live allocations without weakening PTSan's
spatial and temporal protections.

A complementary direction builds on PTSan's optimizer-visible checks
through proof-guided and proof-verified loop specialization.  In this
approach, profiling identifies hot loops, and candidate guards express
the data-structure invariants under which selected checks are
redundant.  CodeHawk-C, our abstract-interpretation
framework~\cite{codehawk}, generates the corresponding memory-safety
proof obligations and verifies that the guards discharge every
obligation associated with an elided check.  Runtime dispatch selects
the specialized loop only when its verified guard holds and otherwise
executes the fully checked version, preserving PTSan's enforcement
while allowing application invariants to remove substantially more
hot-path work.

We are also extending PTSan's response to detected violations.  PTSan
already blocks an invalid access before it executes; that enforcement
point can also serve as a controlled transfer to an appropriate
recovery boundary, such as request cancellation, input rejection, an
existing error path, or worker restart.  Our initial work uses a
violation's call-graph path to locate developer-written error handling
and route control to it automatically.  The longer-term goal is
policy-guided recovery that preserves availability without turning
recovery into silent continuation: every violation remains reported,
the offending operation remains blocked, and recovery proceeds only
through an isolation or error-handling path whose restoration of
program invariants can itself be verified.

\section{Conclusion}
\label{sec:conclusion}

We presented PTSan, an LLVM sanitizer that makes pointer-based memory
safety practical by carrying a compact object identifier in the
pointer itself and keeping bounds and lifetime state in small fixed
tables.  Because identity travels with the pointer value, ordinary
dataflow propagates it for free, and the work that remains at an
access stays ordinary LLVM IR that the compiler can hoist, merge, and
elide.  On SPEC CPU 2017, PTSan adds \SpecOvNolamAll{} runtime
overhead on stock x86-64 hardware, \SpecOvArmAll{} on ARM64, and
\SpecOvLamAll{} with Intel LAM, with
\MemPtsanAll{} memory use, while detecting inter-object and temporal
violations in MSET that location-based sanitizers miss.  These
results show that the long-standing cost gap between location-based
and identity-based sanitizers is not fundamental: when identity
propagation is free and checks are left visible to the optimizer, the
gap closes to parity, with the residual cost of tag stripping reduced
in software by ID-strip hoisting and removed entirely by hardware
address masking.

\section*{Acknowledgment}
This material is based upon work supported by the Defense Advanced
Research Projects Agency (DARPA) under the Enhanced SBOM for Optimized
Software Sustainment (E-BOSS) program, contract HR001124C0486. The
views, opinions, and findings expressed are those of the authors and
should not be interpreted as representing the official views or
policies of the Department of Defense or the U.S.\ Government.

\balance
\printbibliography
\clearpage

\appendices

\section{Additional Evaluation Data}
\label{app:eval}

This appendix collects per-benchmark and supplementary data for the
evaluation of Section~\ref{sec:evaluation} and the related work of
Section~\ref{sec:related-work}.

\begin{figure}[h]
  \centering
  \begin{tikzpicture}
    \begin{axis}[
        width=\columnwidth,
        height=5cm,
        xlabel={Runtime overhead (\%)},
        ylabel={Cumulative \% of programs},
        ylabel near ticks,
        ymin=0, ymax=100,
        xmin=-20, xmax=200,
        tick label style={font=\scriptsize},
        label style={font=\small},
        legend style={font=\scriptsize, draw=none},
        legend pos=south east,
        ymajorgrids,
        const plot,
        no marks,
      ]
      \addplot table [x=ov, y=cum, col sep=comma]
        {eval/generated/llvm-cdf-nolam.csv};
      \addplot table [x=ov, y=cum, col sep=comma]
        {eval/generated/llvm-cdf-lam.csv};
      \addplot table [x=ov, y=cum, col sep=comma]
        {eval/generated/llvm-cdf-arm.csv};
      \legend{PTSan-x86, PTSan-LAM, PTSan-ARM}
    \end{axis}
  \end{tikzpicture}
  \caption{Distribution of PTSan runtime overhead over the measured
  LLVM MultiSource programs (\LlvmMeasuredLam{} on x86-64,
  \LlvmMeasuredArm{} on ARM64).  The axis is clipped at 200\%; two
  ARM64 measurements exceed it (\texttt{sqlite3} at 253\% and
  \texttt{lambda} at 534\%).}
  \label{fig:llvm-cdf}
  \Description{CDF of per-program PTSan runtime overhead on LLVM
  MultiSource benchmarks, on x86-64 with and without LAM and on
  ARM64.}
\end{figure}
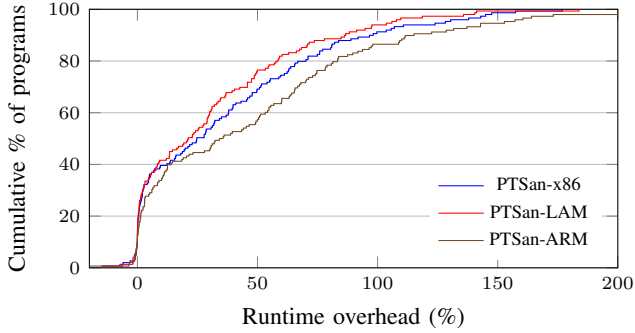

Figure~\ref{fig:llvm-cdf} shows the distribution of per-program
overhead on the LLVM MultiSource benchmarks summarized in
Table~\ref{tab:llvm-summary}.

\begin{table}[h]
  \centering
  \caption{Per-benchmark peak physical memory overhead on SPEC CPU
    2017 (cgroup~v2 \texttt{memory.peak}).}
  \label{tab:spec-memory}
  \footnotesize
  \begin{tabular}{@{}lrrrr@{}}
  \toprule
  Benchmark & Native MiB & ASan & RSan & PTSan \\
  \midrule
  \multicolumn{5}{@{}l}{\emph{SPECrate 2017 Integer}} \\
  500.perlbench\_r & 209 & 4.44$\times$ & 4.54$\times$ & 1.000$\times$ \\
  505.mcf\_r & 616 & 1.47$\times$ & 1.57$\times$ & 1.002$\times$ \\
  520.omnetpp\_r & 262 & 4.45$\times$ & 4.80$\times$ & 0.989$\times$ \\
  523.xalancbmk\_r & 543 & 5.29$\times$ & 4.63$\times$ & 1.001$\times$ \\
  525.x264\_r & 1727 & failed & 1.31$\times$ & 1.015$\times$ \\
  531.deepsjeng\_r & 707 & 1.00$\times$ & 1.04$\times$ & 1.005$\times$ \\
  541.leela\_r & 31 & 34.32$\times$ & 37.18$\times$ & 1.121$\times$ \\
  557.xz\_r & 783 & failed & 1.45$\times$ & 1.003$\times$ \\
  \midrule
  \multicolumn{5}{@{}l}{\emph{SPECrate 2017 Floating Point}} \\
  508.namd\_r & 167 & 2.96$\times$ & 2.89$\times$ & 1.014$\times$ \\
  510.parest\_r & 418 & failed & 4.36$\times$ & 1.007$\times$ \\
  519.lbm\_r & 417 & 1.13$\times$ & 1.07$\times$ & 1.005$\times$ \\
  538.imagick\_r & 291 & 2.75$\times$ & 2.76$\times$ & 1.010$\times$ \\
  544.nab\_r & 152 & 3.50$\times$ & 3.91$\times$ & 1.016$\times$ \\
  \midrule
  \textbf{Geomean} & & \textbf{3.33$\times$} & \textbf{3.01$\times$} & \textbf{1.014$\times$} \\
  Max & & 34.32$\times$ & 37.18$\times$ & 1.121$\times$ \\
  Completed & & 10/13 & 13/13 & 13/13 \\
  \bottomrule
\end{tabular}

  \Description{Table of per-benchmark peak-memory overhead for
    ASan, RSan, and PTSan on SPEC CPU 2017.}
\end{table}

\begin{table}[h]
  \centering
  \caption{MSET detection results for cases with satisfied preconditions.}
  \label{tab:mset-supported}
  \scriptsize
  \resizebox{\linewidth}{!}{%
  \begin{tabular}{@{}lcccccc@{}}
    \toprule
    System & Linear & Non-linear & Type conf. & UAF/UAR & Double-free & Misuse-free \\
    \midrule
    ASan & 46/66 & 0/54 & 6/18 & 8/16 & 4/4 & 20/20 \\
    RSan & 40/72 & 0/54 & 10/24 & 2/8 & 4/4 & 0/20 \\
    PTSan & 54/72 & 36/54 & 12/18 & 16/16 & 4/4 & 20/20 \\
    \bottomrule
  \end{tabular}%
}

  \Description{Table of MSET supported-only detection counts by
    bug family for ASan, RSan, and PTSan.}
\end{table}

\begin{table*}[t]
  \centering
  \caption{Design-space summary of representative sanitizers: SPEC
  geomean overheads and MSET detection coverage.}
  \label{tab:design-space}
  \footnotesize
  \setlength{\tabcolsep}{4pt}
  \begin{tabular}{@{}lllrccl@{}}
    \toprule
    System & Temporal & Runtime & Memory & MSET inter-object & MSET temporal & Deployment model \\
    \midrule
    \multicolumn{7}{@{}l}{\emph{Location-based}} \\
    ASan~\cite{serebryany2012asan} & quarantine & 117\%\textsuperscript{a} & 3.33$\times$\textsuperscript{a} & 90/126 & 32/40 & recompile \\
    RSan~\cite{gorter2025rangesanitizer} & quarantine & 51\%\textsuperscript{b} & 3.01$\times$\textsuperscript{b} & 79/126 & 14/40 & recompile + allocator/loader \\
    Low-Fat~\cite{duck2016lowfat,duck2017stack} & none & 54\%\textsuperscript{c} & 1.0--1.1$\times$\textsuperscript{c} & --- & --- & recompile + allocator \\
    HWASan~\cite{serebryany2018hwasan} & probabilistic & 132--268\%\textsuperscript{c} & 1.04--1.06$\times$\textsuperscript{c} & --- & --- & recompile (Arm TBI) \\
    \midrule
    \multicolumn{7}{@{}l}{\emph{Identity-based}} \\
    SoftBound+CETS~\cite{nagarakatte2009softbound,nagarakatte2010cets} & deterministic & 140--161\%\textsuperscript{c} & 2.7$\times$\textsuperscript{c} & 126/126 & 40/40\textsuperscript{c} & recompile, per-pointer metadata \\
    CUP~\cite{burow2018cup} & probabilistic & 158\%\textsuperscript{c} & n/r & --- & --- & recompile + allocator \\
    Fil-C~\cite{filc-invisicaps} & deterministic & up to $\sim$300\%\textsuperscript{c} & n/r & --- & --- & new ABI + GC runtime \\
    PTSan & probabilistic & \textbf{57\%}/55\%/46\%\textsuperscript{d} & \textbf{1.014$\times$} & \textbf{126/126} & \textbf{40/40} & recompile (LAM: custom kernel) \\
    \bottomrule
  \end{tabular}

  \smallskip
  \begin{minipage}{0.92\textwidth}
    \footnotesize\raggedright
    Memory is peak resident growth over native; MSET columns are the
    inter-object spatial and summed temporal buckets (full
    denominators); ---~not available; n/r~not reported.
    \textsuperscript{a}~Measured locally
    (Section~\ref{sec:evaluation:setup}); ASan runtime is from an
    earlier pass on the PTSan host.
    \textsuperscript{b}~Measured locally against its own baseline on
    a 13-benchmark subset.
    \textsuperscript{c}~As published: Low-Fat and CUP on SPEC CPU
    2006; SoftBound+CETS as the modern port on the SPEC CPU 2017 C
    subset~\cite{10.1145/3642974.3652285}; HWASan as measured on Arm
    TBI and Intel LAM hardware~\cite{gorter2025rangesanitizer}; Fil-C
    as self-reported in its project
    documentation~\cite{filc-invisicaps}.
    \textsuperscript{d}~x86-64 / ARM64 / x86-64 with LAM
    (Section~\ref{sec:evaluation:spec}).
  \end{minipage}
  \Description{Table summarizing temporal mechanism, runtime and
  memory overhead, MSET coverage, and deployment model across
  location-based and identity-based sanitizers.}
\end{table*}

Table~\ref{tab:spec-memory} reports the per-benchmark peak-memory
measurements behind the aggregates of
Section~\ref{sec:evaluation:memory}.  For memory, all systems are
measured against the same native baseline on the same machine; the
mixed-language benchmarks and \texttt{502.gcc\_r} are excluded because
RSan does not cover or build them
(Section~\ref{sec:evaluation:setup}), and ``failed'' marks ASan runs
that crashed before completion.  Table~\ref{tab:mset-supported}
restricts the local MSET runs of Section~\ref{sec:evaluation:mset} to
supported cases only, excluding precondition failures from both
numerator and denominator.  Denominators differ per system because
each sanitizer perturbs allocation behavior differently, so different
tests fail their preconditions under each; SoftBound+CETS is omitted
because only the full-denominator convention is published.

Table~\ref{tab:design-space} summarizes guarantees, costs, and
deployment models across the location-based and identity-based
sanitizers discussed in Sections~\ref{sec:evaluation}
and~\ref{sec:related-work}.

\begin{table*}[t]
  \centering
  \caption{Per-benchmark optimization ablation on SPEC CPU 2017
  (cumulative waterfall): column $-n$ is the configuration after the
  first $n$ removals, in the row order of
  Table~\ref{tab:ablation-summary}.}
  \label{tab:ablation}
  \scriptsize
  \resizebox{\textwidth}{!}{\begin{tabular}{@{}lrrrrrrrrrrr@{}}
  \toprule
  Benchmark & Full & $-$1 & $-$2 & $-$3 & $-$4 & $-$5 & $-$6 & $-$7 & $-$8 & $-$9 & $-$10 \\
  \midrule
  \multicolumn{12}{@{}l}{\emph{SPECrate 2017 Integer}} \\
  500.perlbench\_r & 117.2\% & 181.4\% & 177.5\% & 181.4\% & 186.1\% & 185.9\% & 187.0\% & 197.8\% & 208.1\% & 202.9\% & 246.7\% \\
  502.gcc\_r & 61.4\% & 109.4\% & 113.8\% & 116.5\% & 114.5\% & 115.5\% & 121.7\% & 130.0\% & 134.0\% & 125.7\% & 136.0\% \\
  505.mcf\_r & 32.3\% & 36.2\% & 44.5\% & 42.9\% & 42.8\% & 41.8\% & 43.5\% & 53.4\% & 53.2\% & 54.0\% & 60.9\% \\
  520.omnetpp\_r & 61.8\% & 73.5\% & 74.9\% & 74.3\% & 83.5\% & 84.4\% & 83.5\% & 87.8\% & 89.2\% & 90.7\% & 89.1\% \\
  523.xalancbmk\_r & 57.7\% & 75.8\% & 74.0\% & 75.9\% & 80.3\% & 80.2\% & 80.8\% & 88.5\% & 88.1\% & 92.6\% & 110.4\% \\
  525.x264\_r & 40.6\% & 44.8\% & 56.1\% & 71.9\% & 79.2\% & 88.5\% & 88.4\% & 154.6\% & 159.3\% & 164.9\% & 208.7\% \\
  531.deepsjeng\_r & 49.8\% & 56.9\% & 62.3\% & 66.7\% & 79.9\% & 81.2\% & 81.5\% & 98.8\% & 103.2\% & 103.5\% & 130.9\% \\
  541.leela\_r & 57.3\% & 64.2\% & 71.4\% & 79.8\% & 110.7\% & 111.9\% & 111.5\% & 118.6\% & 121.9\% & 120.6\% & 136.1\% \\
  557.xz\_r & 31.9\% & 34.0\% & 37.4\% & 45.8\% & 47.3\% & 48.3\% & 47.2\% & 52.4\% & 55.4\% & 53.5\% & 68.8\% \\
  \textbf{Geomean (INT)} & \textbf{55.0\%} & \textbf{70.6\%} & \textbf{75.2\%} & \textbf{80.2\%} & \textbf{87.8\%} & \textbf{89.3\%} & \textbf{90.0\%} & \textbf{104.6\%} & \textbf{107.5\%} & \textbf{107.3\%} & \textbf{125.2\%} \\
  \midrule
  \multicolumn{12}{@{}l}{\emph{SPECrate 2017 Floating Point}} \\
  507.cactuBSSN\_r & 57.2\% & 57.2\% & 58.8\% & 60.0\% & 60.1\% & 76.0\% & 76.2\% & 182.4\% & 181.8\% & 237.5\% & 312.0\% \\
  508.namd\_r & 62.9\% & 73.9\% & 80.0\% & 101.1\% & 99.1\% & 116.9\% & 119.4\% & 203.3\% & 194.4\% & 162.0\% & 210.4\% \\
  510.parest\_r & 39.3\% & 39.0\% & 58.5\% & 60.4\% & 59.4\% & 110.5\% & 112.9\% & 130.0\% & 130.3\% & 167.0\% & 191.4\% \\
  511.povray\_r & 118.3\% & 136.5\% & 138.6\% & 148.6\% & 458.6\% & 458.6\% & 459.8\% & 491.8\% & 499.5\% & 491.4\% & 523.1\% \\
  519.lbm\_r & -3.1\% & -2.1\% & -2.9\% & -9.8\% & -9.6\% & -9.4\% & -11.3\% & 33.4\% & 34.6\% & 32.5\% & 88.7\% \\
  526.blender\_r & 8.8\% & 18.0\% & 31.7\% & 34.1\% & 31.9\% & 34.6\% & 34.8\% & 54.1\% & 50.6\% & 56.6\% & 72.8\% \\
  538.imagick\_r & 36.7\% & 45.4\% & 45.8\% & 45.6\% & 75.8\% & 75.7\% & 76.3\% & 78.6\% & 84.2\% & 90.2\% & 93.2\% \\
  544.nab\_r & 12.4\% & 17.9\% & 22.3\% & 23.8\% & 23.8\% & 26.9\% & 27.5\% & -8.0\% & -7.8\% & -9.2\% & 32.8\% \\
  \textbf{Geomean (FP)} & \textbf{37.3\%} & \textbf{43.3\%} & \textbf{49.3\%} & \textbf{51.7\%} & \textbf{71.2\%} & \textbf{82.3\%} & \textbf{82.6\%} & \textbf{111.1\%} & \textbf{111.1\%} & \textbf{117.5\%} & \textbf{157.7\%} \\
  \midrule
  \textbf{Geomean (combined)} & \textbf{46.4\%} & \textbf{57.2\%} & \textbf{62.5\%} & \textbf{66.2\%} & \textbf{79.8\%} & \textbf{86.0\%} & \textbf{86.5\%} & \textbf{107.6\%} & \textbf{109.2\%} & \textbf{112.0\%} & \textbf{139.9\%} \\
  \bottomrule
\end{tabular}
}
  \Description{Table of per-benchmark PTSan overhead as optimizations
  are cumulatively disabled.}
\end{table*}

Table~\ref{tab:ablation} reports the per-benchmark data behind the
ablation summary of Section~\ref{sec:evaluation:ablation}.

\begin{table*}[t]
  \centering
  \caption{Per-benchmark runtime overhead on SPEC CPU 2017 with stack
  temporal protection disabled and enabled.}
  \label{tab:stack-temporal}
  \footnotesize
  \begin{tabular}{@{}cc@{}}
\begin{tabular}{@{}lrrrr@{}}
  \toprule
  \multicolumn{5}{c}{\emph{SPECrate 2017 Integer}} \\
  & \multicolumn{2}{c}{PTSan-LAM} & \multicolumn{2}{c}{PTSan-x86} \\
  Benchmark & Disabled & Enabled & Disabled & Enabled \\
  \midrule
  500.perlbench\_r & 117.1\% & 125.8\% & 179.4\% & 197.1\% \\
  502.gcc\_r & 61.5\% & 65.8\% & 112.2\% & 116.9\% \\
  505.mcf\_r & 32.5\% & 33.1\% & 36.5\% & 35.6\% \\
  520.omnetpp\_r & 62.5\% & 68.5\% & 72.4\% & 79.6\% \\
  523.xalancbmk\_r & 58.9\% & 63.3\% & 76.6\% & 85.9\% \\
  525.x264\_r & 40.5\% & 49.6\% & 44.6\% & 53.4\% \\
  531.deepsjeng\_r & 50.7\% & 56.9\% & 58.6\% & 66.5\% \\
  541.leela\_r & 56.8\% & 70.5\% & 63.9\% & 77.8\% \\
  557.xz\_r & 32.1\% & 32.2\% & 32.2\% & 34.4\% \\
  \textbf{Geomean (INT)} & \textbf{55.3\%} & \textbf{61.0\%} & \textbf{70.6\%} & \textbf{77.8\%} \\
  \midrule
  & & & & \\
  \bottomrule
\end{tabular} &
\begin{tabular}{@{}lrrrr@{}}
  \toprule
  \multicolumn{5}{c}{\emph{SPECrate 2017 Floating Point}} \\
  & \multicolumn{2}{c}{PTSan-LAM} & \multicolumn{2}{c}{PTSan-x86} \\
  Benchmark & Disabled & Enabled & Disabled & Enabled \\
  \midrule
  507.cactuBSSN\_r & 58.1\% & 58.6\% & 60.0\% & 57.7\% \\
  508.namd\_r & 62.8\% & 65.2\% & 73.8\% & 74.5\% \\
  510.parest\_r & 39.2\% & 35.8\% & 38.8\% & 39.7\% \\
  511.povray\_r & 117.3\% & 218.3\% & 133.3\% & 240.1\% \\
  519.lbm\_r & -0.9\% & -1.6\% & -2.1\% & -2.1\% \\
  526.blender\_r & 8.8\% & 10.3\% & 17.8\% & 20.9\% \\
  538.imagick\_r & 36.1\% & 46.0\% & 44.7\% & 54.1\% \\
  544.nab\_r & 12.7\% & -27.3\% & 18.0\% & -21.2\% \\
  & & & & \\
  \textbf{Geomean (FP)} & \textbf{37.6\%} & \textbf{37.9\%} & \textbf{43.2\%} & \textbf{44.3\%} \\
  \midrule
  \textbf{Geomean (combined)} & \textbf{46.7\%} & \textbf{49.7\%} & \textbf{57.1\%} & \textbf{61.2\%} \\
  \bottomrule
\end{tabular} \\
\end{tabular}

  \Description{Table of per-benchmark PTSan runtime overhead with and
  without stack temporal protection under PTSan-LAM and PTSan-x86.}
\end{table*}

Table~\ref{tab:stack-temporal} isolates the runtime cost of stack
temporal protection with and without LAM.

\subsection{Server Workloads}
\label{app:servers}

Table~\ref{tab:servers} gives the full results behind
Section~\ref{sec:evaluation:nginx}.  We ran each Phoronix profile in
its default ``target'' configuration in three configurations (native,
PTSan-x86, and PTSan-LAM).  The workloads are:

\begin{itemize}
  \item \textbf{Redis} (SET, 50 connections): \texttt{redis-benchmark}
  issues SET commands of short string values against the
  single-threaded \texttt{redis-server} over 50 connections.

  \item \textbf{PostgreSQL} (\texttt{pgbench}, buffer test, 1 thread,
  read-write): the TPC-B-like mixed SELECT/UPDATE/INSERT profile with
  one client thread, committing to the write-ahead log on each
  transaction.

  \item \textbf{NGINX} (concurrency 1): \texttt{wrk} fetches a static
  page over HTTPS at one concurrent connection, exercising the TLS
  handshake and request-parsing path, with static delivery through
  \texttt{sendfile}.

  \item \textbf{OpenSSL} (RSA-4096): \texttt{openssl speed rsa4096},
  reporting sign and verify operations per second; asymmetric crypto
  dominated by bignum modular exponentiation.

  \item \textbf{memcached} (Set:Get 1:100): \texttt{memtier\_benchmark}
  drives the in-memory key/value store at a 1:100 Set:Get ratio,
  exercising the slab allocator and hash table.

  \item \textbf{Apache} (100 concurrency): \texttt{ab} serves a static
  page at 100 concurrent requests under the prefork worker model.

  \item \textbf{SQLite} (insert, single thread): replays 2{,}500
  autocommit inserts three times against an on-disk WAL database from
  one thread, timing wall-clock seconds; every commit syncs, making it
  storage-bound.

\end{itemize}

\begin{table*}[t]
  \centering
  \caption{PTSan overhead and peak live-ID demand on Phoronix Test
  Suite server and cryptographic workloads, each in its default
  configuration on the Intel machine of
  Section~\ref{sec:evaluation:setup}.  Overhead is relative to native
  performance; a negative value is faster than native.  SQLite reports
  wall-clock time, so its overhead is added latency.}
  \label{tab:servers}
  \footnotesize
  \setlength{\tabcolsep}{20pt}
  \begin{tabular}{@{}lrrrr@{}}
  \toprule
  Workload & Native & PTSan-x86 & PTSan-LAM & Peak IDs \\
  \midrule
  Redis & 293,701~RPS & 26.0\% & 25.0\% & 16,411 \\
  PostgreSQL & 505~TPS & 16.0\% & 12.0\% & 9,247 \\
  NGINX & 1,802~RPS & 44.0\% & 2.0\% & 7,577 \\
  OpenSSL (sign) & 323~sign/s & 9.0\% & 8.0\% & 7,505 \\
  OpenSSL (verify) & 23,200~verify/s & 3.0\% & 3.0\% & 7,505 \\
  memcached & 524,000~Op/s & 9.0\% & 8.0\% & 706 \\
  Apache & 16,500~RPS & 0.0\% & 0.0\% & 380 \\
  SQLite & 2.5~s & 5.0\% & 6.0\% & 296 \\
  \bottomrule
\end{tabular}

  \Description{Table of per-workload PTSan and PTSan-LAM overhead and
  peak live object-ID counts for eight Phoronix server and
  cryptographic benchmark configurations.}
\end{table*}

\end{document}